\documentclass[camera,letterpaper,nomarginnotes,nonarrowgutter]{jpaper}
\newcounter{version}
\PassOptionsToPackage{hyphens}{url}
\usepackage[bookmarks=true,breaklinks=true,colorlinks,linkcolor=black,citecolor=blue,urlcolor=black]{hyperref}
\usepackage{tikz}
\usepackage{mathptmx} 
\usepackage{amsmath,amssymb,amsfonts}
\usepackage{comment}
\usepackage{algorithm}
\usepackage[noend]{algorithmic}
\usepackage{graphics}
\usepackage{fancyhdr}
\usepackage{booktabs}
\usepackage{pifont}
\usepackage[normalem]{ulem}
\usepackage[]{url}
\usepackage[noadjust]{cite}
\usepackage{hhline}
\usepackage{xcolor}
\usepackage{setspace}
\usepackage{textcomp}
\usepackage{float}
\usepackage[]{enumitem}
\usepackage{afterpage}
\usepackage{graphicx}
\usepackage[]{hyphenat}
\usepackage{makecell}
\usepackage{xspace}
\usepackage{listings}
\usepackage{multirow}
\usepackage{balance}
\usepackage{glossaries} 
\usepackage{siunitx}
\usepackage{duckuments} 
\usepackage{subcaption}
\usepackage{rotating}
\usepackage[us,12hr]{datetime}
\usepackage[en-GB, useregional=numeric]{datetime2}

\usepackage{tablefootnote}

\newcommand{\X}[0]{CoMeT}

\newcommand{\rh}[0]{RowHammer}

\newcommand{\exploitingRowHammerAllCitations}[0]{\cite{fournaris2017exploiting, poddebniak2018attacking, tatar2018throwhammer, carre2018openssl, barenghi2018software, zhang2018triggering, bhattacharya2018advanced, google-project-zero, kim2014flipping, rowhammergithub, seaborn2015exploiting, van2016drammer, gruss2016rowhammer, razavi2016flip, pessl2016drama, xiao2016one, bosman2016dedup, bhattacharya2016curious, burleson2016invited, qiao2016new, brasser2017can, jang2017sgx, aga2017good, mutlu2017rowhammer, tatar2018defeating, gruss2018another, lipp2018nethammer, van2018guardion, frigo2018grand, cojocar2019eccploit,  ji2019pinpoint, mutlu2019rowhammer, hong2019terminal, kwong2020rambleed, frigo2020trrespass, cojocar2020rowhammer, weissman2020jackhammer, zhang2020pthammer, yao2020deephammer, deridder2021smash, hassan2021utrr, jattke2022blacksmith, tol2022toward, kogler2022half, orosa2022spyhammer, zhang2022implicit, liu2022generating, cohen2022hammerscope, zheng2022trojvit, fahr2022frodo, tobah2022spechammer, rakin2022deepsteal}}

\newcommand{\mitigatingRowHammerAllCitations}[0]{\cite{AppleRefInc, rh-hp,rh-lenovo,greenfield2012throttling, kim2014flipping, kim2014architectural, bains14d, bains14c, bains14a, bains14b, aweke2016anvil, bains2015row, bains2016row, bains2016distributed, son2017making, seyedzadeh2018cbt,irazoqui2016mascat, you2019mrloc, lee2019twice, park2020graphene, yaglikci2021security, yaglikci2021blockhammer, frigo2020trrespass, kang2020cattwo, hassan2021utrr, qureshi2022hydra, saileshwar2022randomized, brasser2017can, konoth2018zebram, van2018guardion, vig2018rapid,  kim2022mithril, lee2021cryoguard, marazzi2022protrr, zhang2022softtrr, joardar2022learning, juffinger2023csi, yaglikci2022hira, saxena2022aqua, enomoto2022efficient, manzhosov2022revisiting, ajorpaz2022evax, naseredini2022alarm, joardar2022machine, hassan2022case, zhang2020leveraging,loughlin2021stop, devaux2021method, fakhrzadehgan2022safeguard, saroiu2022price, loughlin2022moesiprime, han2021surround, mutlu2022fundamentally, woo2022scalable, bock2019riprh, kim2015architectural, wang2021discreet, bennett2021panopticon, olgun2024abacus}}
\newcommand{\citeHardWareBasedMitigations}[0]{\cite{kim2014flipping, son2017making, you2019mrloc, lee2019twice, park2020graphene,yaglikci2021blockhammer , greenfield2012throttling, qureshi2022hydra, saileshwar2022randomized,kim2022mithril, yaglikci2022hira, saxena2022aqua, marazzi2022protrr,hassan2022case, devaux2021method, saroiu2022price, han2021surround, mutlu2022fundamentally, woo2022scalable, kim2015architectural, wang2021discreet, bennett2021panopticon, joardar2022learning, joardar2022machine, naseredini2022alarm, zhou2022ltpim, lee2021cryoguard, saroiu2022configure, gomez2016dummy, kang2020cattwo, seyedzadeh2017cbt, seyedzadeh2018cbt, hassan2019crow, ryu2017overcoming, yang2016suppression,olgun2024abacus}}
\newcommand{\citeIntegrityBasedMitigations}{\cite{dell1997white, huang2010ivec, saileshwar2018synergy, chen2014memguard, juffinger2023csi, fakhrzadehgan2022safeguard, qureshi2021rethinking, manzhosov2022revisiting}}
\newcommand{\mitigatingRowHammerCounterCitations}[0]{\cite{kim2014flipping, seyedzadeh2018cbt, park2020graphene, qureshi2022hydra, kim2022mithril, marazzi2022protrr, yaglikci2022hira, loughlin2021stop, kim2014architectural,saxena2022aqua,saileshwar2022randomized}}

\newcommand{\rowHammerGetsWorseCitations}[0]{\cite{kim2014flipping, kim2020revisiting, frigo2020trrespass, yaglikci2022understanding, orosa2021deeper, mutlu2017rowhammer, mutlu2018rowhammer, mutlu2019rowhammer, mutlu2022fundamentally}}

\newcommand{\mitigatingRowHammerCounterBasedIndustry}[0]{\cite{hong2023dsac,kim2023isscc,bennett2021panopticon,devaux2021method, yaglikci2021security,bains14c,bains14d,greenfield2012throttling,greenfield2014method}}
\newcommand{\nrh}[0]{N_{RH}}
\newcommand{\npr}[0]{N_{PR}}

\newcommand{\pprevref}[0]{p_{REF}}
\newcommand{\pnorefinaburst}[0]{p_{B}}
\newcommand{\cboost}[0]{c_B}
\newcommand{\creduc}[0]{c_R}
\newcommand{\pmin}[0]{p_{min}}
\newcommand{\pmax}[0]{p_{max}}

\newcommand{\pnorefuntilxy}[0]{P_{x,y}}
\newcommand{\pij}[0]{p_{i,j}}
\newcommand{\psa}[0]{p_{sa}}
\newcommand{\pfa}[0]{p_{fa}}
\newcommand{\pattack}[0]{p_{attack}}

\newcommand{\psecure}[0]{p_{security}}
\newcommand{\timeout}[0]{t_{to}}

\newcommand{\trcd}[0]{t_{RCD}}
\newcommand{\tras}[0]{t_{RAS}}
\newcommand{\trp}[0]{t_{RP}}
\newcommand{\trc}[0]{t_{RC}}
\newcommand{\trefi}[0]{t_{REFI}}
\newcommand{\trefw}[0]{t_{REFW}}
\newcommand{\trfc}[0]{t_{RFC}}
\newcommand{\trrd}[0]{t_{RRD}}

\newcommand{\act}[0]{ACT}
\newcommand{\pre}[0]{PRE}
\newcommand{\refresh}[0]{REF}
\newcommand{\writecmd}[0]{WR}
\newcommand{\rd}[0]{RD}

\newacronym{iqr}{$IQR$}{inter-quartile range}
\newacronym{act}{$\act{}$}{activate}
\newacronym{pre}{$\pre{}$}{precharge}
\newacronym{ref}{$\refresh{}$}{refresh}
\newacronym{wr}{$\writecmd{}$}{write}
\newacronym{rd}{$\rd{}$}{read}
\newacronym{nrh}{$\nrh$}{RowHammer threshold}
\newacronym{npr}{$\npr$}{preventive refresh threshold}
\newacronym{pprevref}{$\pprevref{}$}{preventive refresh probability}
\newacronym{pnorefinaburst}{$\pnorefinaburst$}{the probability of experiencing no preventive refresh during a burst}
\newacronym{pnorefuntilxy}{$\pnorefuntilxy$}{the probability of experiencing \emph{no} preventive refresh until a given $x_{th}$ row activation in $y_{th}$ burst of activations}
\newacronym{cboost}{$\cboost{}$}{the coefficient of boosting preventive refresh probability}
\newacronym{creduc}{$\creduc{}$}{the coefficient of reducing preventive refresh probability}
\newacronym{pij}{$\pij{}$}{the probability of experiencing no preventive refresh after the row activation $i$ at a given batch $j$}
\newacronym{psa}{$\psa{}$}{the probability of a successful attempt}
\newacronym{pfa}{$\pfa{}$}{the probability of a failed attempt}
\newacronym{pattack}{$\pattack{}$}{the probability of a successful attack}
\newacronym{psecure}{$\psecure{}$}{\gls{mech}'s security guarantee over a time period, T,}
\newacronym{trefw}{$\trefw$}{refresh window}
\newacronym{trcd}{$\trcd$}{{trcd definition here}}
\newacronym{tras}{$\tras$}{the latency of fully restoring a DRAM cell's charge}
\newacronym{trp}{$\trp$}{{trp definition here}}
\newacronym{trc}{$\trc$}{the minimum time needed between two consecutive row activations targeting the same bank}
\newacronym{trefi}{$\trefi$}{refresh interval}
\newacronym{trfc}{$\trfc$}{refresh latency}
\newacronym{trrd}{$\trrd$}{the minimum time needed between two consecutive row activations targeting the same rank}
\newacronym{timeout}{$\timeout$}{timeout period}
\newacronym{pmin}{$\pmin$}{{DEFINE PMIN PLEASE}}
\newacronym{pmax}{$\pmax$}{{DEFINE PMAX PLEASE}}

\newcommand{\secref}[1]{§\ref{#1}}
\newcommand{\figref}[1]{Fig.~\ref{#1}}

\usepackage{tikz}

\newcommand*\circled[1]{\tikz[baseline=(char.base)]{\node[shape=circle,fill,inner sep=0.5pt] (char) {\textcolor{white}{#1}};}}

\newcommand{\head}[1]{{\noindent\textbf{#1.}\xspace}}

\definecolor{nbs}{rgb}{0.88, 0.07, 0.37}
\definecolor{iy}{rgb}{0.0, 0.36, 0.05}
\definecolor{gfored}{rgb}{0.580, 0.050, 0.211}

\definecolor{revc}{rgb}{0.19, 0.55, 0.91}
\definecolor{revd}{rgb}{0.87, 0.36, 0.51}
\definecolor{reve}{rgb}{1.0, 0.5, 0.31}

\definecolor{revb}{rgb}{0.9, 0.17, 0.31}
\definecolor{cq}{rgb}{0.0, 0.5, 1.0}

\definecolor{bluepigment}{rgb}{0.2, 0.2, 0.6}
\definecolor{brandeisblue}{rgb}{0.0, 0.44, 1.0}

\newif\ifdraft
\draftfalse

\ifdraft
    \usepackage[colorinlistoftodos,prependcaption,textsize=small]{todonotes}
    \paperwidth=\dimexpr \paperwidth + 4cm\relax
    \oddsidemargin=\dimexpr\oddsidemargin + 2cm\relax
    \evensidemargin=\dimexpr\evensidemargin + 2cm\relax
    \marginparwidth=\dimexpr \marginparwidth + 2cm\relax
    \newcommand{\nbcomment}[1]{\todo[size=\scriptsize, linecolor=nbs, bordercolor=nbs, backgroundcolor=white]{\textcolor{nbs}{\textbf{@nb:} #1}}}
    \newcommand\param[1]{{\color{blue}{#1}}}
    \newcommand\nb[1]{{\color{nbs}{#1}}}
    \definecolor{yt}{rgb}{0.58, 0.44, 0.86}
    
    \newcommand{\yctcomment}[1]{\textcolor{yt}{\textbf{[@yct:} #1\textbf{]}}}

    \newcommand{\atbcomment}[1]{\todo[size=\scriptsize, linecolor=orange, bordercolor=orange, backgroundcolor=white]{\textcolor{orange}{\textbf{@atb:} #1}}}
    \newcommand\atb[1]{{\color{orange}{#1}}}

    \newcommand{\agycomment}[1]{\todo[size=\scriptsize, linecolor=purple, bordercolor=purple, backgroundcolor=white]{\textcolor{purple}{\textbf{@agy:} #1}}}
    \newcommand{\ieycomment}[1]{\todo[size=\scriptsize, linecolor=orange, bordercolor=orange, backgroundcolor=white]{\textcolor{iy}{\textbf{@iey:} #1}}}

\else 
    \definecolor{shadecolor}{RGB}{180,180,180}
    \definecolor{darkgray}{rgb}{0.86, 0.86, 0.86}
    \usepackage[framemethod=tikz]{mdframed}
    \newcommand{\fancycommand}[1]{%
    \begin{mdframed}[backgroundcolor=darkgray, linecolor=black, linewidth=0.5pt]
        \texttt{#1}%
    \end{mdframed}%
}

    \newcommand{\revb}[1]{}

    \newcommand{\revc}[1]{}

    \newcommand{\revdt}[1]{{#1}}
    \newcommand{\revd}[1]{}

    \newcommand{\revet}[1]{{#1}}
    \newcommand{\reve}[1]{}
     
     \newcommand{\cq}[1]{}

    \newcommand{\nbcomment}[1]{}
    \newcommand{\atbcomment}[1]{}
    
    \newcommand\param[1]{{\color{black}{#1}}}
    \newcommand\nb[1]{#1}

    \newcommand\atb[1]{#1}

    \newcommand{\yctcomment}[1]{}
    
    \newcommand{\ieycomment}[1]{}
    
    \newcommand{\agycomment}[1]{}
\fi

\newif\ifcameraready
\camerareadytrue

\ifcameraready
  \newcommand\nbcr[2]{#2}
  \newcommand\atbcr[2]{#2}

\else 
     \newcommand\nbcr[2]{\ifnum#1=\value{version}\textcolor{red}{#2}\else{\textcolor{blue}{#2}}\fi}
    \newcommand\atbcr[2]{\ifnum#1=\value{version}\textcolor{orange}{#2}\else{\textcolor{blue}{#2}}\fi}

    \usepackage[colorinlistoftodos,prependcaption,textsize=small]{todonotes}
    \usepackage{marginnote} 
    \let\marginpar\marginnote
    \renewcommand{\nbcomment}[1]{\todo[size=\scriptsize, linecolor=nbs, bordercolor=nbs, backgroundcolor=white]{\textcolor{nbs}{\textbf{@nb:} #1}}}
\fi

\newcommand{\versionnum}[0]{3.7}

\ifcameraready
    \fancypagestyle{firstpage}{
        \fancyhead{} 
        \fancyhead[C]{
        \begin{tikzpicture}[remember picture,overlay]
        \node [xshift=153mm,yshift=-10mm]
        at (current page.north west) {{\includegraphics[width=1.8cm]{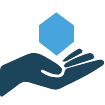}}} ;
        \node [xshift=172mm,yshift=-10mm]
        at (current page.north west) {{\includegraphics[width=1.8cm]{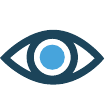}}} ;
        \node [xshift=191mm,yshift=-10mm]
        at (current page.north west) {{\includegraphics[width=1.8cm]{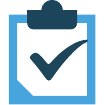}}} ;
        \end{tikzpicture}
      } 
    
    }
\else
    \fancypagestyle{firstpage}
    {
        \fancyhead{}
        \fancyhead[C]{
        \textcolor{red}{CONFIDENTIAL DRAFT -- DO NOT DISTRIBUTE -- TO APPEAR IN HPCA'24} \\ \textcolor{MidnightBlue}{\emph{Version \versionnum~---~\today, \ampmtime}}  \hrule 
        }
    }
\fi

\begin{document}
\bstctlcite{IEEEexample:BSTcontrol}

\title{\X{}: Count-Min\nbcr{1}{-}Sketch-based Row Tracking\\to Mitigate RowHammer at Low Cost}


\hyphenation{}

\newcommand{\hpcapubid}{0000--0000/00\$00.00}
\newcommand\hpcaauthors{First Author$\dagger$ and Second Author$\ddagger$}
\newcommand\hpcaaffiliation{First Affiliation$\dagger$, Second Affiliation$\ddagger$}
\newcommand\hpcaemail{Email(s)}

\author{{F. Nisa Bostanc\i}\qquad%
{İsmail Emir Y\"{u}ksel}\qquad
{Ataberk Olgun}\qquad
{Konstantinos Kanellopoulos}\qquad \\
{Yahya Can Tu\u{g}rul}\qquad
{A. Giray Ya\u{g}l{\i}k\c{c}{\i}}\qquad%
{Mohammad Sadrosadati}\qquad
{Onur Mutlu}\qquad\vspace{-3mm}\\\\
{{ETH Z{\"u}rich}} }


\ifcameraready
  \thispagestyle{firstpage}
\else
  \thispagestyle{firstpage}
\fi

\maketitle
\thispagestyle{firstpage}
\begin{abstract}

DRAM chips are increasingly more vulnerable to read-disturbance phenomena (e.g., RowHammer and RowPress), where repeatedly accessing DRAM rows causes bitflips in nearby rows due to DRAM density scaling. 
\nb{Under low \rh{} thresholds, existing \rh{} mitigations either incur high area overheads or degrade performance significantly. }

We propose a new \rh{} mitigation mechanism, \X{}, that prevents \rh{} bitflips with \nbcr{2}{low area, performance, and energy costs} in DRAM-based systems at \nbcr{1}{very low} \rh{} thresholds.
The key idea of \X{} is to use low-cost and scalable 
hash-based counters to track DRAM row activations\atb{.} 
\nb{\X{} uses the Count-Min Sketch \nbcr{3}{technique} \atbcr{2}{that} 
\nbcr{2}{maps each DRAM row to a group of counters\atbcr{2}{, as uniquely as possible,} using \nbcr{3}{multiple} hash functions. When a DRAM row is activated, \X{} \atbcr{2}{increments \nbcr{2}{the counters mapped to that DRAM row}}.
\atbcr{2}{Because the mapping from DRAM rows to counters is not completely unique,
activating one row can increment one or \nbcr{3}{more} counters mapped to another row.}}
\atbcr{2}{\nbcr{2}{Thus}, \X{} may overestimate, but \emph{never} underestimates, \nbcr{2}{a DRAM row's} activation count. This property of \X{} allows it to securely prevent RowHammer bitflips while \nbcr{2}{properly configuring} its hash functions reduces \nbcr{2}{overestimations}. \nbcr{3}{As a result}, \X{} 1) implements substantially fewer counters (e.g., thousands of counters) than the number of DRAM rows in a DRAM bank (e.g., 128K rows) and 2) does not significantly overestimate \nbcr{2}{a DRAM row's activation count}. \nbcr{3}{We demonstrate that} \X{} securely prevents RowHammer bitflips at low area, performance, and energy cost.}}

Our comprehensive evaluations show that \X{} prevents \rh{} bitflips with an average performance overhead of \textit{only} \param{$0.19$}\% and \param{$4.01$}\%  across 61 benign \nbcr{3}{single-core} workloads \nb{for \nbcr{1}{a} \rh{} threshold of 1K and} \nb{a \nbcr{1}{very} low} \rh{} threshold of 125, respectively, normalized to a system with no \rh{} mitigation.
\X{} achieves a good trade-off between performance, energy, and area overheads. \nb{Compared to the \nbcr{2}{best prior performance- and energy-efficient \rh{} mitigation mechanism}, \X{} requires \param{$5.4\times$} and \param{$74.2\times$} less area overhead at \nbcr{1}{\rh{}} thresholds of 1K and 125, respectively, and incurs {a small} (\param{$\leq1.75$}\%) performance overhead on average, for all \rh{} thresholds.}~Compared to the \nbcr{2}{best prior low-area-cost mitigation mechanism}, at \nbcr{1}{a very} low \rh{} threshold of 125, \X{} \nbcr{3}{improves} performance \nbcr{1}{by} up to $39.1\%$ \nbcr{3}{while incurring a similar area overhead}.
\nbcr{2}{\X{} is openly \nbcr{5}{and freely} available at \url{https://github.com/CMU-SAFARI/CoMeT}.}

\end{abstract}
\glsresetall{}

%

\section{Introduction}

\atb{DRAM chips are susceptible to read-disturbance where repeatedly accessing} \atb{a DRAM row} (i.e., \textit{an aggressor row}) can cause bitflips in physically nearby rows (i.e., \textit{victim rows}) ~\cite{kim2014flipping, mutlu2017rowhammer, yang2019trap, mutlu2019rowhammer,park2016statistical, park2016experiments,
walker2021ondramrowhammer, ryu2017overcoming, yang2016suppression, yang2017scanning, gautam2019row, jiang2021quantifying,luo2023rowpress}. \rh{} is a type of read-disturbance phenomenon that is caused by repeatedly opening and closing  (i.e., \textit{hammering}) DRAM rows. Modern DRAM chips become more vulnerable to \rh{} as DRAM technology \nbcr{1}{node size becomes smaller}~\rowHammerGetsWorseCitations{}\nbcr{1}{:} the minimum number of row activations needed to cause a bitflip (i.e., \textit{\gls{nrh}}) has reduced by more than an order of magnitude in less than a decade~\cite{kim2020revisiting}.\footnote{For example, \gls{nrh} is \emph{only} 4.8K for newer DRAM chips (from 2019--2020), which is $14.4\times$ lower than the \gls{nrh} of 69.2K for some older DRAM chips (from 2010--2013)~\cite{kim2020revisiting}.\label{rh_scale_fn}} \nbcr{2}{Exacerbating} the situation, 
RowPress~\cite{luo2023rowpress} \nbcr{1}{is another widespread read-disturbance phenomenon in} modern DRAM chips \nbcr{1}{that} 
induces bitflips in DRAM rows by keeping them open for excessive periods of time\nbcr{1}{. RowPress} \nbcr{2}{leads to bitflips at} substantially \nbcr{2}{lower} \nbcr{1}{DRAM} row activation \nbcr{3}{counts}.\footnote{RowPress is shown to \nbcr{2}{lead to bitflips} with one to two orders of magnitude fewer activations \nbcr{2}{(than \rh{})} under realistic conditions~\cite{luo2023rowpress}.}
Prior works show that \atb{attackers can leverage} \rh{} bitflips in real systems~\exploitingRowHammerAllCitations{} to\nbcr{1}{, for example,} (i)~take over systems by \nbcr{1}{escalating privilege} and (ii) leak security-critical and private data. Consequently, prior works propose many mitigations to prevent \rh{} bitflips~\mitigatingRowHammerAllCitations{}. 

\head{Key Problem} Many prior works \nbcr{1}{from academia}~\mitigatingRowHammerCounterCitations{} \nbcr{1}{and industry~\mitigatingRowHammerCounterBasedIndustry{}} propose using counter\nbcr{2}{s} to track the activation count\nbcr{2}{s} of potential aggressor row\nbcr{2}{s}. \atb{We refer to these works as} counter-based mechanisms. \nbcr{2}{T}hese mechanisms prevent \rh{} bitflips \nbcr{2}{at} low performance and energy overheads \nbcr{2}{by taking preventive actions precisely when it is required.}
\atb{However, t}hese \atb{mechanisms} have high area overheads 
due to two reasons: \atb{(1)}
\nbcr{2}{tag-based counters are implemented using hardware structures with high area costs, and (2)~the number of counters grows with the increasing number of possible aggressor rows.}

\head{(1)~Expensive hardware structures}
Many mitigations rely on area-hungry \nbcr{2}{hardware} structures~\cite{yaglikci2021blockhammer,park2020graphene,lee2019twice,seyedzadeh2018cbt, kim2022mithril} (e.g., content-addressable memor\nbcr{1}{ies}) to store activation counters. Simply implementing one activation counter for each DRAM row in main memory is very costly~\cite{qureshi2022hydra,kim2014flipping} as doing so requires {MBs} of memory (e.g., 20 MiB for \nbcr{2}{a modern DDR5}~\cite{jedec2020ddr5} \nbcr{2}{channel with 2 ranks and $2^{23}$ rows with 10-bit counters per row}). {To avoid implementing per-row counters,} prior works leverage the key {observation} that only a subset of DRAM rows can be activated \gls{nrh} times in a refresh window.\footnote{\nbcr{2}{Refresh window is} \SI{64}{\milli\second} \nbcr{1}{and \SI{32}{\milli\second}} in DDR4~\cite{jedec2017ddr4} \nbcr{1}{and DDR5~\cite{jedec2020ddr5}, respectively}. \nbcr{2}{When} all DRAM rows are refreshed, the disturbance effects of RowHammer on the victim rows are reset~\cite{kim2014flipping}.} Thus, the number of \emph{possible aggressor rows} is smaller than all DRAM rows in main memory. Therefore, prior works implement \nbcr{1}{fewer} activation counters \nbcr{2}{than} one for each DRAM row\nbcr{2}{,} to \nbcr{1}{track only the most activated rows (i.e., possible aggressor rows).}
To do so, these works 
use content-addressable memor\nbcr{1}{ies} (CAM\nbcr{1}{s}) \nbcr{2}{since} \nbcr{1}{a} CAM allows rapid access to an activation counter of a DRAM row given that DRAM row's address.
{Unfortunately, \nbcr{1}{a} CAM fundamentally takes up more area than other types of memory (e.g., SRAM) due to its large cell size~\cite{mohammad2006cache,yaglikci2021blockhammer}.}

\head{(2)~Large number of counters}
\nbcr{1}{As} \rh{} threshold \nbcr{1}{reduces with technology scaling}, an attacker \nbcr{1}{can} {hammer} more DRAM rows concurrently. 
\nbcr{2}{The area overhead of DRAM row activation counters does not scale well because, at lower \gls{nrh} values, counter-based mechanisms need to allocate more counters to keep track of all possible aggressor rows.}
For example, Graphene's~\cite{park2020graphene} \nbcr{2}{storage} overhead can be as high as \nbcr{1}{$\sim$1.5 MiB} at \gls{nrh}=\nbcr{1}{125} (\secref{sec:motivation}).

Due to \nbcr{2}{these} \nbcr{1}{two} \atb{reasons}, \atb{some} prior works (e.g.,~\cite{qureshi2022hydra, you2019mrloc, son2017making, wang2021discreet, yaglikci2022hira,kim2014flipping, marazzi2023rega}) employ different techniques \nbcr{2}{than storing a large number of counters in the memory controller} to mitigate \rh{} at low area overhead. \atb{For example, Hydra~\cite{qureshi2022hydra} reduce\nbcr{1}{s}} the area overhead \nbcr{2}{in processor} by \atb{partially or completely storing row activation counters in main memory}, and \atb{\nbcr{1}{PARA~\cite{kim2014flipping} and HiRA~\cite{yaglikci2022hira}} randomly identif\nbcr{2}{y} a subset of all activated DRAM rows as possible aggressor rows without using any activation counters.} 
Unfortunately, these works cause prohibitively large performance overheads 
\nbcr{2}{at low} \nbcr{1}{\gls{nrh} values }
\nb{by incurring \nbcr{2}{either (i)} unnecessary preventive refreshes, \nbcr{2}{or (ii)} additional memory requests that take up DRAM bandwidth}.
Therefore, it is important to \atb{design} \rh{} \atb{mitigations} \nbcr{2}{that have} low area and performance overhead\nbcr{1}{s} \atb{as} DRAM chips become more vulnerable \nbcr{1}{to \rh{}}.

\textbf{Our goal} is to design a \rh{} mitigation \nbcr{1}{technique} that prevents \rh{} bitflips with low area, performance, and energy overhead\nbcr{1}{s} in highly \rh{}-vulnerable DRAM-based systems. To this end, we propose \X{}, a \textbf{Co}unt-\textbf{M}in Sketch-based Row \textbf{T}racking \nbcr{3}{Technique} to Mitigate RowHammer at Low Cost. The \textbf{key idea} of \X{} is to use low-cost and scalable hash-based counters to track DRAM rows and \nbcr{1}{thus,} reduce the overhead of expensive tag-based counters. To do so, \X{} leverages \nbcr{1}{the} Count-Min Sketch (CMS) \nbcr{1}{\nbcr{3}{technique}~\cite{cormode2005improved}} 
\nbcr{2}{that maps each DRAM row to a group of counters, as uniquely as possible, using \nbcr{3}{multiple} hash functions. \nbcr{3}{When} a DRAM row is activated, \X{} increments the counters mapped to that DRAM row. \nbcr{3}{Because the mapping from DRAM rows to counters is not} completely unique, \nbcr{3}{activating one row can increment one or more counters mapped to another row. Thus,} \X{} \nbcr{3}{may} overestimate \nbcr{3}{a DRAM row's} activation count. \nbcr{3}{However, \X{} never underestimates a DRAM row's activation count because when a DRAM row is activated, \X{} always increments all corresponding counters.} \nbcr{3}{As a result,} CMS (1) enables securely preventing \rh{} bitflips with substantially fewer counters than the number of DRAM rows, and (2) does not significantly overestimate \nbcr{3}{a} DRAM row's activation count when it is \nbcr{3}{properly configured}.}

\head{Key Mechanism} 
\X{} consists of two main components: \textit{Counter Table} (CT) and \textit{Recent Aggressor Table} (RAT). \textit{Counter Table} accurately tracks DRAM rows' activation counts with a low area overhead by using tagless hash-based counters and employing the CMS \nbcr{3}{technique}. CT maps \nbcr{2}{each DRAM row} to a group of counters
\nbcr{2}{and when a DRAM row is activated, it increments the corresponding counters.}
CT \textit{only} \nbcr{2}{triggers} a preventive refresh when all counters associated with a row exceed a predetermined threshold (which we set to be smaller than \gls{nrh}).
This way, \nbcr{2}{CT tracks DRAM row activation counts with low area overhead and high accuracy.} 
\nbcr{2}{\textit{Recent Aggressor Table} (RAT)} tracks a small set of recently identified aggressor rows with per-row counters.
When a DRAM row's CT counters reach the activation threshold, \nbcr{2}{\X{} does \textit{not} reset the corresponding CT counters because any counter can be shared across different DRAM rows. Instead, \X{} (1) performs preventive refresh and (2)} allocates a RAT entry \nbcr{2}{for the activated row.} The next time the row is activated, only the RAT counter is used to estimate its activation count. \nbcr{2}{Doing so} prevents \X{} from using the saturated CT counters and \nbcr{1}{thus,} \nbcr{2}{performing} \nbcr{1}{unnecessary} preventive refreshes \nbcr{1}{due to \nbcr{2}{the same} aggressor row}.
\nbcr{2}{By combining these two components, \X{} securely prevents \rh{} bitflips at low \rh{} thresholds with low area, performance, and energy costs.}

\head{Hardware Implementation}
\nbcr{1}{We configure CT and RAT to minimize \X{}'s false positives (i.e., overestimations of a DRAM row's activation count) at very low \rh{} thresholds. We evaluate \X{}'s storage and area overhead using CACTI~\cite{cacti}. \nbcr{4}{We model \X{}’s hardware design (RTL) in Verilog and evaluate its circuit area and latency overheads using modern ASIC design tools. We} report that \X{} incurs (i) 76.5 KiB and 51.0 KiB storage overheads and (ii) $0.09mm^2$ and $0.07mm^2$ area overheads at \nbcr{2}{an} \gls{nrh} of 1K and 125, respectively.}

\head{Key Results} We evaluate \X{}'s impact on system performance and energy consumption \nbcr{1}{using Ramulator~\cite{kim2016ramulator, ramulatorgithub}} \nbcr{2}{across} a diverse set of \param{61} single-core and \param{56} multi-core workloads from SPEC CPU2006, SPEC CPU2017, {TPC, MediaBench, and YCSB} benchmark suites. We compare \X{} to four \nbcr{2}{state-of-the-art} \nbcr{1}{\rh{}} mitigations \nbcr{2}{(Graphene~\cite{park2020graphene}, Hydra~\cite{qureshi2022hydra}, REGA~\cite{marazzi2023rega}, PARA~\cite{kim2014flipping})}. We \nbcr{1}{report} four key results.
First, at a \rh{} threshold of 1K, \X{} incurs only \nbcr{1}{(i)}~\param{$0.19$}\% average performance and \param{$0.08$}\% average DRAM energy overheads across single-core workloads 
\nbcr{1}{and (ii)~\param{$0.73$}\% average performance and \param{$0.15$}\% average DRAM energy overheads across 8-core workload mixes\nbcr{2}{,} compared to a system without any \rh{} protection.}
Second, \X{} scales well into future {for DRAM chips with} \nbcr{1}{very} low RowHammer thresholds:\nbcr{2}{~at an \gls{nrh} of 125, \X{} incurs \param{$4.01$}\% average performance and \param{$2.07$}\% average DRAM energy overheads across single-core workloads.}
Third, compared to the \nbcr{2}{best prior performance- and energy-efficient \rh{} mitigation} \nbcr{1}{technique\nbcr{2}{, Graphene}~\cite{park2020graphene}}, \X{} requires \param{$5.4\times$} and \param{$74.2\times$} less area overhead at \gls{nrh}\nbcr{2}{=}1K and 125, respectively, and incurs {a small} (\param{$\leq1.75$}\%) performance overhead on average, at all \rh{} thresholds. \nbcr{1}{Fourth, c}ompared to the \nbcr{2}{best prior low-area-cost} \rh{} mitigation \nbcr{1}{technique\nbcr{2}{, Hydra}~\cite{qureshi2022hydra}, at \gls{nrh}\nbcr{2}{=}$125$}\nbcr{2}{,} \X{} \nbcr{1}{improves performance \nbcr{2}{by} up to \param{$39.1$}\% while incurring \nbcr{3}{a similar} area overhead.} 
\nbcr{2}{We open-source our simulation infrastructure and all datasets at \url{https://github.com/CMU-SAFARI/CoMeT} to enable reproducibility and \nbcr{3}{aid} future research.}

We make the following key contributions:
\sloppy\ignorespaces
\begin{itemize}
\item We introduce \X{}, a new low-cost \rh{} mitigation mechanism \nbcr{3}{that uses the Count-Min Sketch technique~\cite{cormode2005improved} to track DRAM row activations}. \X{} \nbcr{1}{securely} prevents \rh{} bitflips at \nbcr{1}{very} low \rh{} thresholds with low area, performance, and energy overheads, compared to the state-of-the-art \rh{} mitigation mechanisms.
\item We evaluate the performance, energy, and area overheads of four state-of-the-art mechanisms\nbcr{3}{, demonstrating that} \nbcr{2}{(1)} \X{} performs similarly to the \nbcr{2}{best prior performance- and energy-efficient} \rh{} mitigation technique while incurring significantly \nbcr{1}{lower} area overhead ($5.4\times$ and $74.2\times$ less for \gls{nrh}$=$1K and 125)\nbcr{2}{, and (2) \X{} improves performance (by up to \param{39.1}\% at \gls{nrh}$=125$) compared to the best prior low-area-cost \rh{} mitigation technique with similar area overhead.}
\end{itemize}

\glsresetall{}
\setcounter{version}{5}
\section{Background}

\subsection{DRAM Organization and Operation}

\begin{figure}[ht]
\centering
\includegraphics[width=0.9\linewidth]{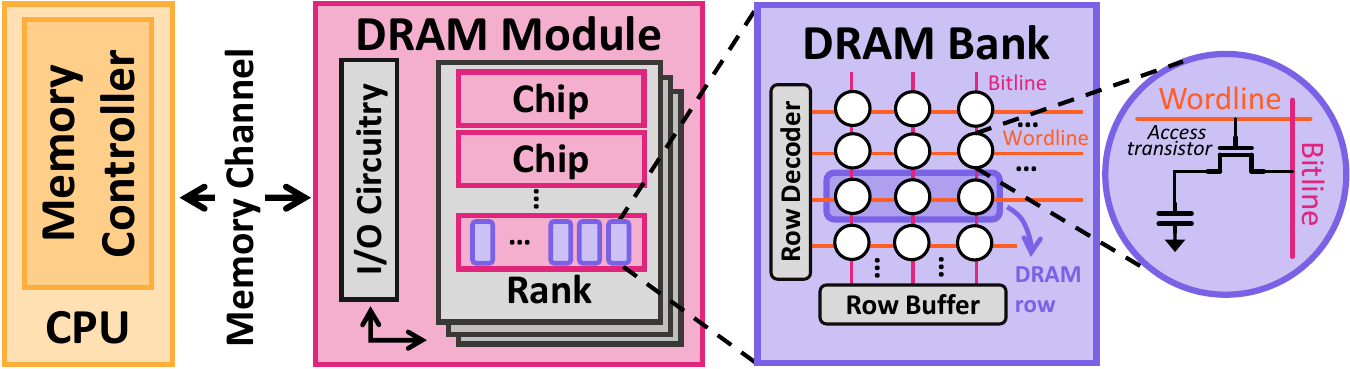}
\caption{DRAM organization.}
\label{fig:dram-organization}
\end{figure}

\head{Organization}~\figref{fig:dram-organization} shows the hierarchical organization of a modern DRAM-based main memory. The memory controller connects to a DRAM module over a memory channel. A module contains one or \nbcr{3}{more} DRAM ranks that time-share the memory channel. A rank consists of multiple DRAM chips that operate in lock-step. 
Each DRAM chip contains multiple DRAM banks that can be accessed independently. A DRAM bank is organized as a two-dimensional array of DRAM cells, where a row of cells is called a DRAM row. A DRAM cell consists of 1) a storage capacitor, which stores one bit of information in the form of electrical charge, and 2) an access transistor, which connects the capacitor to the row buffer through a bitline controlled by a wordline.

\head{Operation}
To access a DRAM row, the memory controller issues a set of commands to DRAM over the memory channel. The memory controller sends an $ACT$ command to activate a DRAM row, which asserts the corresponding wordline and loads the row data into the row buffer. Then, the memory controller can issue $RD$/$WR$ command\nbcr{3}{s} to read from/write into the DRAM row. 
Subsequent accesses to the same row cause a row hit. To access a different row, the memory controller must first close the bank by issuing a $PRE$ command.
Therefore, accessing a different row causes a row miss/conflict.

DRAM cells are inherently leaky and lose their charge over time due to charge leakage in the access transistor and the storage capacitor. \nbcr{3}{T}o maintain data integrity, the memory controller periodically refreshes each row in a time interval called \gls{trefw} \nbcr{4}{which is typically} $32 ms$ for DDR5~\cite{jedec2020ddr5} and $64 ms$ for DDR4~\cite{jedec2017ddr4} \nbcr{4}{at normal operating temperature (i.e., up to \SI{85}{\celsius}) and half of it for the extended temperature range (i.e., above \SI{85}{\celsius} up to \SI{95}{\celsius})}. To ensure all rows are refreshed every \gls{trefw}, the memory controller issues REF commands with a time interval called \gls{trefi} ($3.9 \mu s$ for DDR5~\cite{jedec2020ddr5} and $7.8 \mu s$ for DDR4~\cite{jedec2017ddr4} \nbcr{4}{at normal operating temperature \nbcr{5}{range}}).

\head{Timing Parameters}
To ensure correct operation, the memory controller must obey specific timing parameters while accessing DRAM. In addition to \gls{trefw} and \gls{trefi}, we explain \param{three} timing parameters related to the rest of the paper: i) the minimum time interval between two consecutive ACT commands targeting the same bank ($t_{RC}$), ii) the minimum time needed to issue a PRE command following an ACT command ($t_{RAS}$), and iii) the minimum time needed to issue an ACT command following a PRE command ($t_{RP}$).

\subsection{\rh{} \nbcr{1}{Mitigations}}
\nbcr{1}{To securely prevent \rh{} bitflips and protect DRAM-based computing systems,} prior works propose \nbcr{1}{different} mitigation techniques~\mitigatingRowHammerAllCitations{}. These works \nbcr{1}{take \nbcr{3}{a} preventive action (\nbcr{2}{e.g.,} preventive\nbcr{3}{ly refresh} victim \nbcr{2}{rows} or throttl\nbcr{3}{e} potential aggressor rows) either (i)~\nbcr{3}{by probabilistically sampling DRAM row activations} or (ii) \nbcr{3}{by deterministically tracking the number of times DRAM rows are activated (i.e., row activation counts)}. }\nbcr{1}{As DRAM chips become more vulnerable to RowHammer}, both approaches incur \nbcr{3}{larger} performance, energy, and area overheads. \nbcr{1}{Probabilistic} \nbcr{3}{sampling-based techniques} \nbcr{1}{prevent \rh{} bitflips at low area cost. However, at \nbcr{3}{very low} \gls{nrh} values (e.g., sub-1K), they} suffer from high performance, and energy overheads~\cite{kim2020revisiting, yaglikci2022hira}. Tracking-based \nbcr{3}{techniques} either prevent bitflips (i) with low performance cost but \nbcr{3}{with} high area overheads or (ii) at low area cost but \nbcr{3}{with prohibitively large performance and energy overheads}.

\subsection{\nbcr{1}{Count-Min Sketch \nbcr{3}{Data Structure}}}
\label{sec:cms-background}
\atbcr{3}{Keeping precise track of the number of times each unique item appears (i.e., the \emph{frequency} of an item) in a data stream requires as many counters as there are unique items in the stream. Implementing a counter in hardware requires storage and induces chip area overhead.} \nbcr{3}{Count-Min Sketch (CMS) is a data structure that summarizes a data stream at a smaller storage cost compared to \atbcr{3}{implementing a} counter for \atbcr{3}{each unique} item in the stream. CMS can be used to answer queries such as how many times a specific item \atbcr{3}{appears} in the data stream and provides highly accurate estimations. Thus, \atbcr{3}{CMS} can be used to determine the frequent items in a data stream (i.e., frequent item counting) \atbcr{3}{at low chip area overhead}.}

\begin{figure}[ht]
\centering
\includegraphics[width=\linewidth]{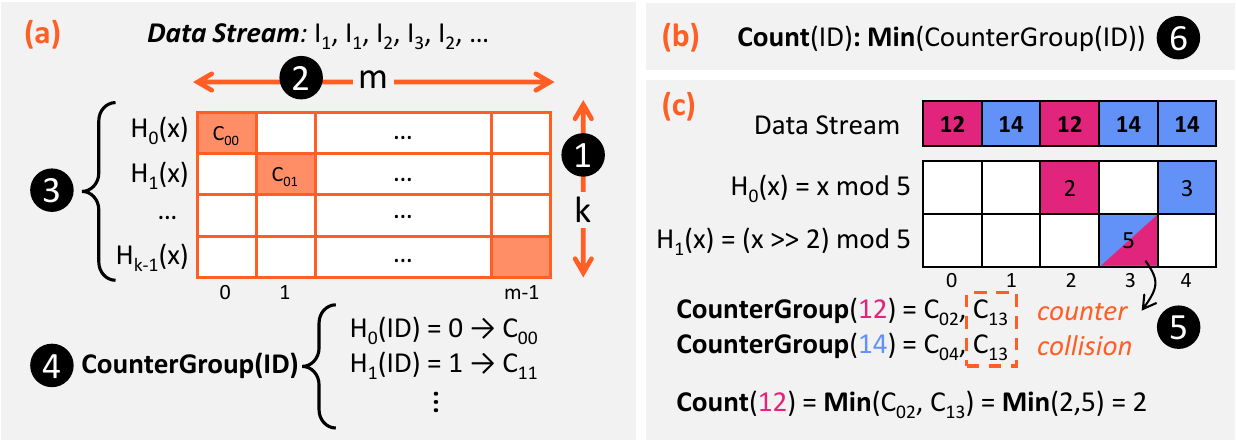}
\caption{Count-Min Sketch overview.}
\label{fig:comet-cms}
\end{figure}

\figref{fig:comet-cms}~\nbcr{3}{(a)} show\nbcr{4}{s} an overview of CMS. CMS uses a hash-based approach and introduces a two-dimensional counter array of size  $(k~rows) * (m~columns)$  (\circled{1} and \circled{2}) that is accessed via \textit{k} hash functions \circled{3}. Each item in the stream is \nbcr{3}{given an item ID and} mapped to a counter group in this two-dimensional counter array \nbcr{3}{by calculating the counter indices using the item ID (\circled{4}).}

\nbcr{3}{\figref{fig:comet-cms}(c) depicts an example with a CMS array of 2 rows and 5 columns \atbcr{3}{(middle)} summarizing the data stream \atbcr{3}{(top)}. CMS uses the following two hash functions to map items to counters: (1) $H_{0}(x)= x$ mod 5, and (2) $H_{1}(x)= (x >> 2)$ mod 5. The first item in the stream has the item ID 12. To determine the counter group for item 12, we calculate the counter indices using \atbcr{3}{the two} hash functions. First, by calculating $H_{1}(12)= 12$ mod 5 $=2$, CMS maps item 12 to $c_{02}$ \atbcr{3}{(column 2 in row 0)}. Second, by calculating $H_{1}(x)= (x >> 2)$ mod 5 $=3$, CMS maps item 12 to $c_{13}$ \atbcr{3}{(column 3 in row 1)}. Similarly, the second item (item ID 14) is mapped to $c_{04}$ and $c_{13}$. }

\head{Updating the counters}
\nbcr{3}{When an item \atbcr{3}{appears} in the stream, CMS increments all counters in \atbcr{3}{the item's} counter group. When there is a collision in one counter across different items' counter groups, such as the \atbcr{3}{one in} counter $c_{13}$ (\circled{5}), CMS increments the colliding counter whenever any item mapped to the same counter occurs in the stream. By choosing a distinct set of hash functions and configur\atbcr{3}{ing} the $k$ and $m$ parameters, CMS reduces the collision rate. }

\head{Estimating the frequency of items in the stream} \nbcr{3}{To estimate the frequency of an item} in the stream, CMS uses the \textit{minimum} counter value in the counter group \nbcr{3}{corresponding to the item}~\circled{6}. CMS can \nbcr{3}{overestimate the \atbcr{3}{frequency} of} an item due to collisions in its group's minimum counter. However, CMS cannot \nbcr{3}{underestimate the frequency of an item} because an item's counters are always incremented when \atbcr{3}{the} \nbcr{3}{item} \atbcr{3}{appears} in the stream and \atbcr{3}{the counters} \nbcr{3}{are} never reset~\cite{cormode2005improved}. This way, CMS provides an upper bound for the \atbcr{3}{frequency} of the item in the stream.

\head{Optimizations} Several works propose optimizations for CMS to reduce its \nbcr{3}{overestimations due to counter collisions}. \nbcr{3}{CMS with Conservative Updates (CMS-CU)~\cite{estan2002new,cohen2003spectral}} limits the number of counter\nbcr{2}{s that are} increment\nbcr{2}{ed during an update} to \nbcr{2}{avoid} \nbcr{3}{overestimations}. When an item occurs in the stream, \nbcr{3}{CMS-}CU \textit{only} increments the \nbcr{2}{counter\nbcr{4}{(s)} with the minimum value}. \nbcr{3}{This way, CMS-CU avoids \atbcr{3}{unnecessarily} incrementing the counters with higher values and prevents overestimations based on these counters in the future.}
Since the minimum counter \nbcr{2}{value} is prove\nbcr{3}{n} to be \atbcr{3}{larger than the actual frequency of an item in the data stream}~\cite{cormode2005improved} and is \nbcr{3}{always incremented \atbcr{3}{with each CMS} update}, this optimization does \textit{not} cause \nbcr{3}{underestimations~\cite{estan2002new,cohen2003spectral}.}

\section{Motivation}
\label{sec:motivation}

\subsection{High Read-Disturbance Vulnerability}
As DRAM chips become more vulnerable to \rh{}, fewer row activations can induce bitflips. 
\nbcr{3}{Thus, an attacker can hammer more rows concurrently in a refresh period}. 
\nbcr{2}{As a result}, \nbcr{3}{at very low \rh{} thresholds,} state-of-the-art RowHammer mitigation mechanisms \nbcr{3}{either 1)} require \nbcr{3}{prohobitively large numbers of} activation counters \nbcr{3}{to track all possible aggressor rows}\nbcr{3}{, or 2) incur high performance and energy overheads due to occupying a significant portion of the DRAM bandwidth with preventive actions (\secref{sec:limits_mitig})}. 
To make matters worse, another \nbcr{3}{widespread} read-disturbance phenomenon, RowPress~\cite{luo2023rowpress}, affect\nbcr{2}{s} modern DRAM chips. RowPress induces bitflips in DRAM rows by keeping physically adjacent rows open for extended periods of time. \nbcr{2}{RowPress leads to bitflips with substantially lower row activation} \nbcr{3}{counts than \rh{}} (e.g., one to two orders of magnitude fewer activations under realistic conditions)~\cite{luo2023rowpress}.
\nbcr{3}{\nbcr{4}{The} RowPress \nbcr{4}{work}~\cite{luo2023rowpress} shows that existing \rh{} mitigation techniques can be adapted to prevent RowPress bitflips by (i)~\nbcr{4}{limiting the time DRAM rows can remain open and (ii)} performing preventive actions at \nbcr{4}{smaller} activation counts \nbcr{4}{corresponding to row open times}. Thus, adapting existing mitigation techniques to take RowPress into account requires secure and performance-, energy-, and area-efficient operation at \nbcr{4}{even lower activation counts}.}

\subsection{Limitations of Existing Mitigations}
\label{sec:limits_mitig}

\head{Performance-Optimized Mitigations}
The \atbcr{3}{most straightforward} way to prevent \rh{} bitflips is to keep track of \atbcr{3}{\emph{all}} DRAM rows' activation counts with \atbcr{3}{a} dedicated counter \atbcr{3}{for each} DRAM row~\cite{kim2014flipping}. \atbcr{3}{Unfortunately}, this method leads to \atbcr{3}{very} large area overheads in systems with high-density DRAM modules. For example, \atbcr{3}{10-bit} counters for a modern \atbcr{3}{DDR5} channel with \atbcr{3}{2 ranks and $2^{23}$} rows~\cite{jedec2020ddr5} would require \atbcr{3}{20} MiB storage to employ \nbcr{1}{per-DRAM-row} counters. 
To overcome this issue, prior works~\cite{park2020graphene,kim2022mithril,saxena2022aqua,saileshwar2022randomized,marazzi2022protrr,olgun2024abacus} employ a frequent item counting algorithm, Misra-Gries~\cite{misra1982finding}, to track a relatively smaller number of DRAM rows that can \atbcr{3}{potentially} reach the \rh{} threshold. However, \nbcr{1}{as DRAM becomes more vulnerable to read-disturb\atbcr{3}{ance, fewer hammers can induce bitflips.}} 
\atbcr{3}{Even though the number of activate and precharge commands that the memory controller can issue in a refresh window remains the same, more rows can be concurrently hammered {\gls{nrh} times at a smaller \gls{nrh}}} (e.g., 2720 and 21760 \atbcr{3}{rows can be hammered} per bank at \gls{nrh}$=1K$ and \gls{nrh}$=125$\nbcr{4}{, respectively}).
Consequently, counter-based techniques \atbcr{3}{(\nbcr{4}{e.g., those} that use Misra-Gries)} require \nbcr{1}{significantly larger} number\nbcr{1}{s} of counters \atbcr{3}{as DRAM scales down to smaller technology nodes.} 

To demonstrate this \atbcr{3}{problem}, we use a 
\atbcr{3}{state-of-the-art Misra-Gries-based} \rh{} mitigation mechanism \atbcr{3}{that has very low system performance and DRAM energy overheads}, Graphene~\cite{park2020graphene}, as a concrete example \nbcr{1}{(implemented as described in~\cite{park2020graphene} (\secref{sec:methodology}))}. 
Table~\ref{table:graphene_motiv} shows Graphene's 32-bank \atbcr{3}{(2-rank)} storage overhead for different \gls{nrh} values. We observe that with \nbcr{2}{lower} \gls{nrh} \nbcr{2}{values}, the \nbcr{1}{storage} cost of Graphene increases significantly. \atbcr{3}{Graphene's storage cost increases from $207.2$ KiB at \gls{nrh}$=$1K to {$\sim1.5$ MiB} at \gls{nrh}$=$125.}

\begin{table}[ht]
\centering
\caption{Storage overhead of \atbcr{3}{a} performance-optimized state-of-the-art \rh{} mitigation~\cite{park2020graphene}.}
\label{table:graphene_motiv}
\resizebox{0.7\linewidth}{!}{
\begin{tabular}{l|llll}
\textbf{$N_{RH}$} & \textbf{1000} & \textbf{500} & \textbf{250} & \textbf{125}  \\ 
\hline\hline
\textbf{Storage (KB)}  & \nbcr{1}{207.19}        & \nbcr{1}{498.44}       & \nbcr{1}{765.00}         & \nbcr{1}{1466.25}         
\end{tabular}
}
\end{table}

\head{Area-Optimized Mitigations}
\nbcr{2}{Several prior works~\cite{qureshi2022hydra, you2019mrloc, son2017making, wang2021discreet, yaglikci2022hira,kim2014flipping, marazzi2023rega} aim {to prevent} RowHammer bitflips at \atbcr{3}{extremely} low area costs \nbcr{4}{by employing different techniques}.} \nbcr{4}{Best performing low-area-cost \rh{} mitigation mechanism, Hydra~\cite{qureshi2022hydra},} tracks row activations with per-DRAM-row activation counters located in main memory (i.e., DRAM chip). To minimize DRAM access overhead during activation\nbcr{2}{s}, Hydra employs a \nbcr{4}{filtering} logic that groups rows into fixed DRAM row groups, and each group is assigned a \textit{group counter}. Initially, DRAM row activations only increment the group counters, and only when a group counter reaches a predetermined threshold, the group counter value is copied to the per-DRAM-row counters in DRAM. After this, only the per-row counters are incremented upon DRAM row activations, and the group counter is not used until the end of the refresh period. \nbcr{4}{Hydra caches per-row counters inside the memory controller to avoid incurring frequent memory accesses.} Using this technique, Hydra is able to prevent \rh{} bitflips with a small processor chip area overhead.\footnote{Other mitigation techniques that employ probabilistic methods or modify DRAM design incur higher performance overheads than Hydra at low \gls{nrh}, as shown in~\secref{sec:results}.}

\nbcr{4}{However, Hydra incurs a high performance overhead at low \gls{nrh}, due to two key drawbacks. First, Hydra \textit{overestimates} the activation counts of DRAM rows and performs many unnecessary preventive refreshes. Hydra's group counters can reach the group counter threshold quickly when a memory-intensive application activates many DRAM rows in the same group only a few times. As a result, Hydra overestimates the activation counts and preventively refreshes many rows even though no DRAM row is activated enough times to cause a bitflip.
Second, Hydra occupies a significant portion of the DRAM bandwidth to access and write to per-DRAM-row counters in DRAM. When a per-DRAM-row activation counter is not cached in the memory controller, Hydra needs to fetch the counter from the main memory and place it in the cache. Doing so incurs additional memory latency for fetching the new counter and writing back an evicted counter. }

\begin{figure}[ht]
\centering
\includegraphics[width=0.7\linewidth]{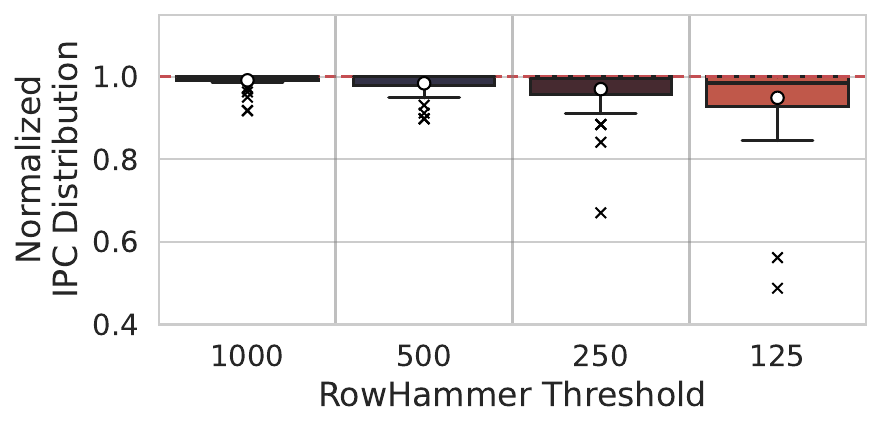}
\caption{Performance overhead of \atbcr{3}{an} area-optimized state-of-the-art \rh{} mitigation mechanism~\cite{qureshi2022hydra}.}
\label{fig:hydra_performance}
\end{figure}

We model Hydra (explained in detail in~\secref{sec:methodology}) and measure its performance impact on 61 benign single-core applications across \atbcr{3}{various} \gls{nrh} values.~\figref{fig:hydra_performance} shows the IPC distribution of \nbcr{5}{61} benign single-core applications normalized to a baseline with no \rh{} mitigation as a box plot.\footnote{\label{footnote:boxplot}{Each box in the figure represents the interquartile range of the observed IPC values. The midlines of the boxes indicate the median value of the corresponding interquartile ranges. Cross marks \nbcr{3}{($\times$)} show the outlier values.}} 
We \nbcr{4}{make two observations. First,} \atbcr{3}{as \gls{nrh} \nbcr{5}{reduces} from 1K to 125}, Hydra's performance overhead increase\nbcr{2}{s} significantly. Hydra has an average (maximum) performance overhead of $0.85\%$ ($8.18\%$) at \gls{nrh}$=1K$. The average performance overhead increases to $5.66\%$ ($51.24\%$) at \gls{nrh}\nbcr{1}{$=$}125. \atbcr{3}{These overheads increase to 6.11\% and 25.02\% in multi-core (8-core) workloads.}
Second, \nbcr{4}{Hydra increases memory read latency \nbcr{5}{significantly} due to \nbcr{5}{its} (i) preventive refreshes and (ii) \nbcr{5}{off-chip memory accesses to fetch per-DRAM-row activation counters. At \gls{nrh}$=125$, Hydra increases} average memory read latency by 16.91\% (5.36$\times$) compared to the baseline system across all single-core (multicore) workloads.} 

\subsection{Our Goal}

\begin{figure}[ht]
\centering
\includegraphics[width=\linewidth]{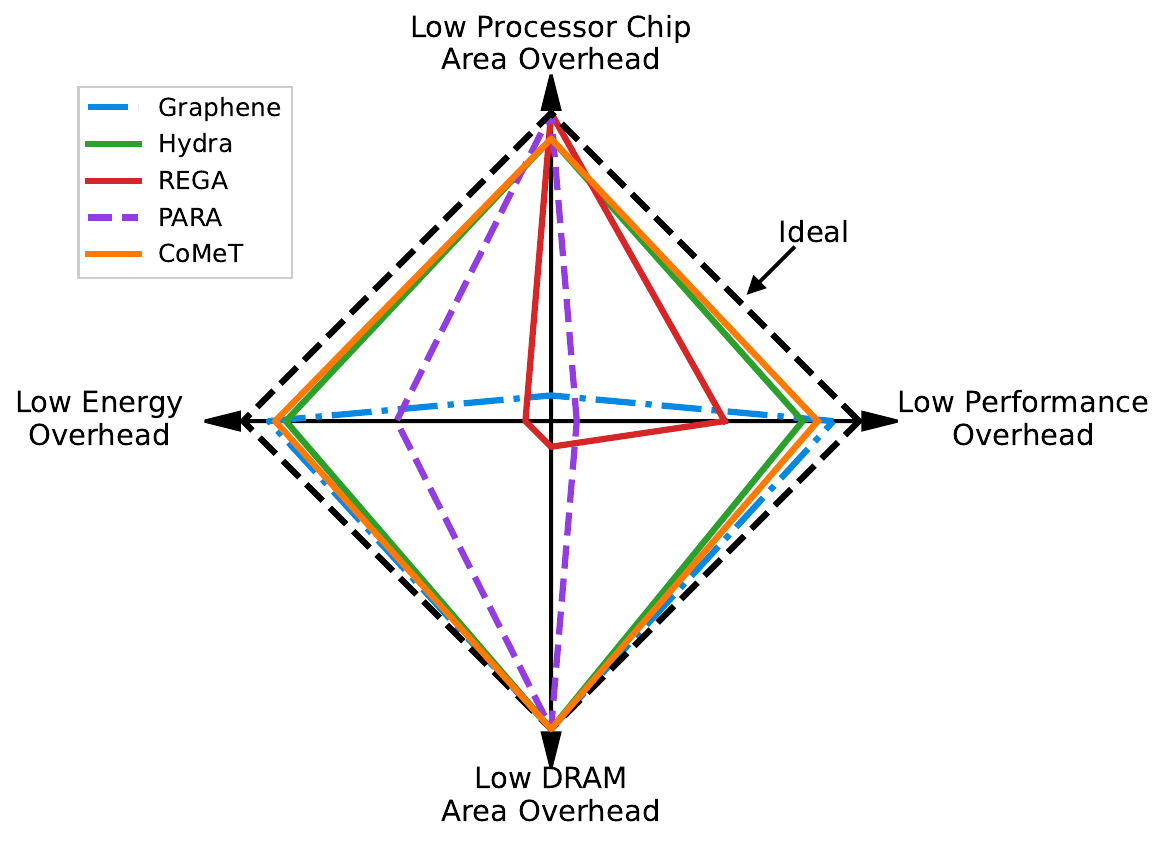}
\caption{Trade-off between performance, \nbcr{4}{processor area, DRAM area,} and energy costs of existing \rh{} mitigation \nbcr{1}{mechanisms and \X{}}.
}
\label{fig:tradeoff}
\end{figure}

\nb{To summarize our observations,} {\figref{fig:tradeoff} compares four state-of-the-art \rh{} mitigations at \nbcr{3}{\gls{nrh}$=$125}, using a \nbcr{5}{four}-dimensional radar plot with performance, \nbcr{4}{processor chip} and DRAM chip area overheads, and energy consumption on \nbcr{5}{four} different axes.\footnote{Results across 61 single-core workloads at \gls{nrh}$=125$. \nbcr{5}{\secref{sec:methodology} describes our methodology and \secref{sec:results} provides all our results.}}
We plot the negative of the \nbcr{5}{processor chip and DRAM chip} area overhead, the negative of the \nbcr{4}{average} performance overhead, and the negative of the \nbcr{4}{average} energy overhead on the respective axes.\footnote{\nbcr{3}{Note that we assume PARA and REGA's \nbcr{5}{processor chip} area overheads are the smallest as they do not incur any significant area overhead on processor chip: (i)~PARA is stateless and does not store any metadata, and \nbcr{4}{(ii)} REGA modifies the DRAM chip design and only incurs a DRAM area overhead.}} The ideal \rh{} mitigation should have \nbcr{4}{zero} area, performance, and energy \nbcr{3}{overheads} and is plotted as the black dashed line in the figure.}  

\nb{We observe that no existing \rh{} mitigation achieves a balance between low area and performance overheads while also maintaining low energy consumption. PARA \nbcr{1}{ha\nbcr{5}{s} negligible} \nbcr{4}{processor} \nbcr{5}{and DRAM} \nbcr{3}{chip} area overheads, but \nbcr{1}{also incur\nbcr{5}{s} high performance and energy overheads at low \gls{nrh} values}. \nbcr{5}{REGA has a negligible processor chip area overhead but has a fixed DRAM area overhead and incurs high performance and energy overheads.} \nbcr{1}{Similarly,} Hydra \nbcr{1}{prevents \rh{} bitflips with a low} \nbcr{4}{processor \nbcr{5}{and DRAM} chip} area overhead\nbcr{5}{s}, but it does so at the cost of high performance overhead and energy consumption.\footnote{\nbcr{4}{Hydra implements per-DRAM-row counters in DRAM (in addition to its SRAM-based counters) and incurs a storage overhead of 4 MiB for 8-bit counters.\label{footnotehydra}}}
\nbcr{1}{In contrast with PARA, REGA, and Hydra,} Graphene \nbcr{1}{prevents \rh{} bitflips at} low performance and energy \nbcr{3}{overheads} yet incurs \nbcr{3}{very} high \nbcr{5}{processor chip} area overhead. }

\nbcr{3}{As DRAM-based systems become more vulnerable to read-disturbance phenomena, it is critical to prevent \rh{} bitflips at very low \gls{nrh} values.}
\nbcr{3}{Unfortunately, as \gls{nrh} reduces, existing \rh{} mitigation techniques incur either }
\nbcr{1}{1)} high area overheads due to \nbcr{1}{tracking large} number\nbcr{1}{s} of aggressor rows or \nbcr{1}{2)} high performance and energy overheads due to occupying increasingly more DRAM bandwidth \nbcr{1}{with excessive numbers of} preventive refreshes and \nbcr{3}{main} memory accesses. 
\textbf{Our goal} is to design a \rh{} mitigation \nbcr{1}{technique} that prevents \rh{} bitflips with low area, performance, and energy \nbcr{1}{overheads} in \nbcr{1}{highly \rh{}-vulnerable} DRAM-based systems.

\setcounter{version}{6}
\section{\X{}}
\head{Overview}
\nbcr{3}{\X{} is designed to securely prevent \rh{} bitflips with low area, performance, and energy overheads at low \gls{nrh} values. Achieving this goal requires accurately tracking DRAM row activation counts at low area cost and preventively refreshing victim rows only when it is necessary. To this end, \X{} employs the Count-Min \nbcr{4}{S}ketch (CMS) \nbcr{5}{technique}~\cite{cormode2005improved} (\secref{sec:cms-background}) to track DRAM row activations due to two main reasons. First, CMS enables tracking DRAM row activations at low area cost by (1) implementing substantially fewer counters than the number of DRAM rows and (2) using low-cost counters \nbcr{6}{accessed with hash functions (i.e., \textit{hash-based counters})} that fundamentally take up less area \nbcr{4}{than counters \nbcr{6}{accessed with tag-matching (i.e., \textit{tag-based counters})}}. Second, CMS does not significantly overestimate DRAM row activation counts when configured properly (\secref{sec:exploration}). This way, \X{} \nbcr{4}{avoid\nbcr{5}{s} performing many} preventive \nbcr{5}{refreshes} due to overestimations. }

\nbcr{5}{\X{} employs the CMS technique \nbcr{6}{to} map \nbcr{6}{each} DRAM row to a group of counters, as uniquely as possible, using multiple hash functions and performs a preventive refresh when a DRAM row's counter group reaches a predetermined activation threshold (i.e., \textit{\gls{npr}}). A row that reaches \gls{npr} is called an \textit{aggressor row}. \nbcr{6}{Ideally, after preventively refreshing an aggressor row's victims, activation counters of the aggressor row can be reset. However, because}
CMS's mapping from DRAM rows to counters is not completely unique, 
resetting the counters of an aggressor row 
can result in underestimating the activation count of another DRAM row mapped to the same counter(s).} 
\nbcr{5}{To ensure no DRAM row's activation count is underestimated, \X{} does \textit{not} reset any hash-based counter after a preventive refresh. As a result, a hash-based counter that reaches \gls{npr} \textit{is saturated} (i.e., its value stays at \gls{npr}). A saturated counter overestimates the activation count of the aggressor row that caused the preventive refresh and continues to trigger \textit{unnecessary} preventive refreshes. Therefore, adopting CMS is not sufficient to achieve high performance and energy efficiency.}

To reduce the unnecessary preventive refreshes, \X{} allocates \nbcr{4}{a small \nbcr{6}{table} of} \atbcr{5}{tag-based} per-DRAM-row counters for \nbcr{5}{DRAM rows that are estimated to be activated \gls{npr} times by hash-based counters (\atbcr{5}{we call such aggressor rows} \textit{recent aggressor rows}).
When a DRAM row is activated, \X{} checks whether a dedicated per-DRAM-row counter is allocated for it. \X{} uses the per-DRAM-row counter to estimate the row's activation count and increments it \nbcr{6}{only} when the same row is activated. T}his way, \X{} accurately \nbcr{5}{estimates the activation count of a DRAM row with \atbcr{5}{a} per-DRAM-row counter \textit{after} its victims are preventively refreshed and \nbcr{6}{thus,} avoids \nbcr{6}{further} unnecessary preventive refreshes \nbcr{6}{for this row}.} \nbcr{4}{\X{} combines hash-based and tag-based counters to accurately track DRAM row activations and prevents \rh{} bitflips at low performance and energy overheads. By allocating per-DRAM-row counters to \textit{only} a small set of DRAM rows, \X{} incurs a low area overhead.}

\begin{figure}[ht]
\centering
\includegraphics[width=\linewidth]{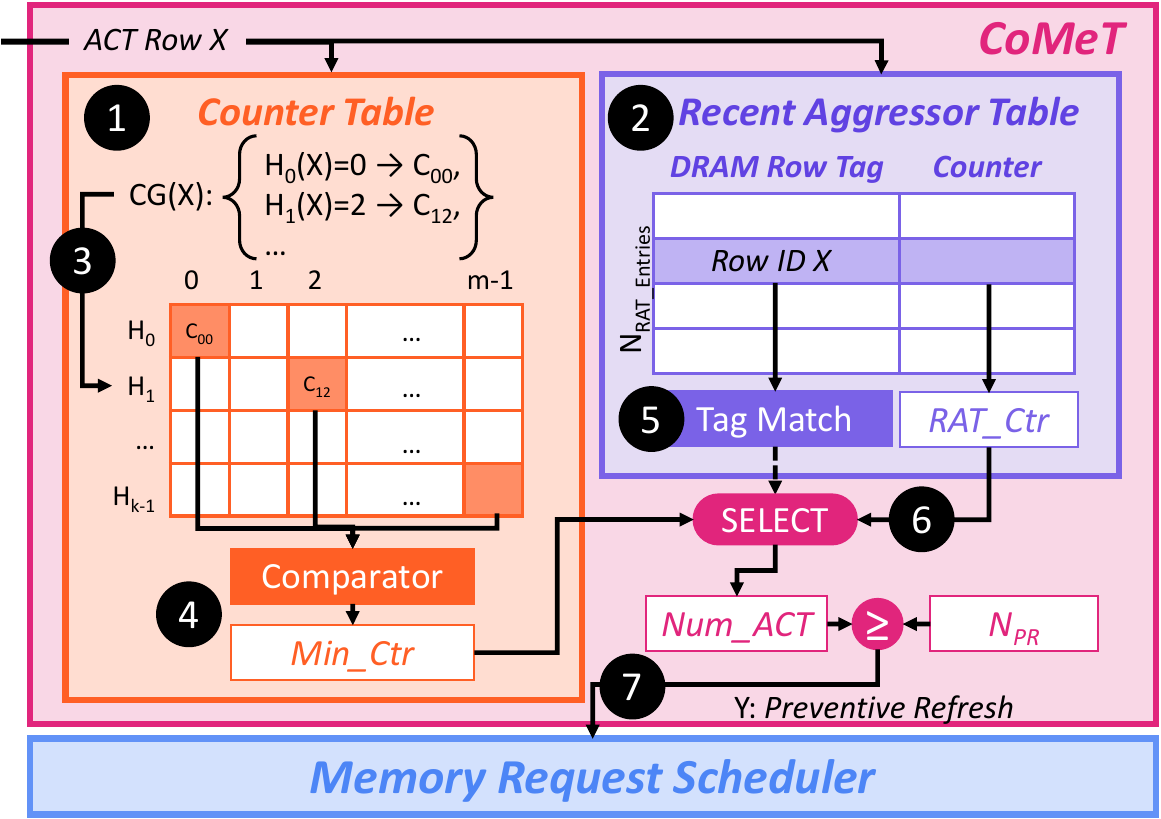}
\caption{\X{} Overview.}
\label{fig:comet-overview}
\end{figure}

\head{Key Components}
~\figref{fig:comet-overview} presents an overview of \X{} \nbcr{3}{and its two key components: \textit{Counter Table} \circled{1} that consists of hash-based activation counters tracking all DRAM rows, and \textit{Recent Aggressor Table} \circled{2} that consists of per-DRAM-row \nbcr{4}{activation} counters tracking \nbcr{4}{only} \nbcr{4}{a small set of DRAM rows that reach \gls{npr} activations}.}
\nbcr{3}{\textbf{Counter Table} (CT) aims to track row activation counts with high accuracy and low area overhead. To do so, CT (i) interprets the row activation tracking problem as a frequent item counting problem and (ii) employs the Count-Min Sketch \nbcr{5}{technique} with conservative updates (\secref{sec:cms-background}). CT consists of $m$ counters for each of $k$ hash functions \atbcr{5}{(i.e., \X{} uses each of the $k$ hash functions to index a different set of $m$ counters)} per bank and uses simple hash functions that consist of bit-shift and bit-mask operations, which are easy to implement in hardware.}
\nbcr{3}{\textbf{Recent \nbcr{4}{Aggressor} Table} (RAT) aims to prevent unnecessary preventive refreshes caused by saturated CT counters. RAT consists of $N_{RAT\_Entries}$ per-DRAM-row counters \atbcr{5}{each of which is tagged} with \nbcr{4}{a DRAM row ID (i.e., DRAM row address bits)}. A RAT counter is allocated \nbcr{4}{\textit{only}} when a DRAM row reaches \gls{npr} activations \atbcr{5}{(in the CT)} and \atbcr{5}{the RAT counter} always stores the actual activation count of that DRAM row. We explain how we determine the sizes of each key component in \secref{sec:exploration}.}

\subsection{Operation of \X{}} 
\nbcr{3}{We describe \X{}'s operation in five steps: (i) initialization, (ii) activation count estimation, (iii) updating \X{}'s counters, (iv) early \nbcr{6}{preventive} refresh, and (v) periodic reset.}

\head{1. Initialization}
\nbcr{3}{Initially and after early \nbcr{6}{preventive} refresh (Step 4) and periodic reset (Step 5), no DRAM rows are activated \nbcr{5}{after their victim rows are refreshed}. Therefore, \nbcr{5}{the disturbance effects of \rh{} are reset~\cite{kim2014flipping} and} all CT and RAT counters are set to zero. }

\head{2. Activation Count Estimation}
The memory controller issues an $ACT$ command to a row ID \texttt{X} in a bank. Consequently, \X{} concurrently accesses \nbcr{4}{CT and RAT} with the row ID. 

\head{2.a. Accessing CT} { 
\nbcr{4}{\X{} determines the counter group of row X (\textit{CounterGroup(X)}) (\circled{3}) by (i) calculating the result of each hash function with the row ID X (e.g., $H_{0}(X)=0$ and $H_{1}(X)=2$), and (ii) using each hash function\nbcr{5}{'s output} to access a counter across different CT rows  (e.g., CT row 0: $C_{00}$ and CT row 1: $C_{12}$).}
Then, CT estimates row \texttt{X}'s activation count as the minimum counter value, \textit{Min\_Ctr}, across all counters in the counter group \nbcr{4}{(e.g., minimum value across ($C_{00}$,$C_{12}$,...)) (\circled{4}).}

\head{2.b. RAT Search} {Concurrent with the CT access, \X{} searches row {ID} \texttt{X} in the RAT tag array to determine if a per-DRAM-row counter is allocated for the row. If there is a tag match, RAT estimates the activation count as the corresponding counter value, \textit{RAT\_Ctr} \nbcr{4}{(\circled{5})}.}

\noindent\X{} estimates the activation count \nbcr{4}{(\textit{Num\_ACT})} as \textit{RAT\_Ctr} if there is a tag-match in RAT, otherwise as \textit{Min\_Ctr} (\circled{6}).

\head{3. Updating \X{}'s Counters} \X{} calculates the \nbcr{4}{updated} activation count \nbcr{4}{(i.e., the activation count including the current activation)} \nbcr{6}{by incrementing \textit{Num\_ACT} by one} and compares the \nbcr{4}{updated} activation count with \gls{npr}. 

\nbcr{3}{If the activation count is \nbcr{5}{greater than or} equal to \gls{npr}, \nbcr{4}{i.e., \textit{Num\_ACT}$\geq$\gls{npr},} \X{} preventively refreshes the victim rows \nbcr{4}{(\circled{7})}. Then, \nbcr{4}{\X{} sets the counters in \textit{CounterGroup(X)} to \gls{npr}. If there is a RAT entry already allocated for row X, \X{} sets the existing RAT counter to zero. Otherwise, \X{} allocates a new RAT entry for row X. When RAT is fully occupied by other rows,} \X{} randomly selects a RAT entry to evict. Note that \X{} ensures that this eviction is safe because evicted rows continue to use their CT counters, which are {already} at \gls{npr}. \nbcr{4}{However, \nbcr{5}{RAT evictions can} result in unnecessary preventive refreshes due to overestimati\nbcr{5}{on of} the \nbcr{5}{evicted} row's activation count.} }

\nbcr{3}{If the activation count of the row is smaller than \gls{npr}, \nbcr{4}{i.e., \textit{Num\_ACT}$<$\gls{npr}}, \X{} only updates the row's counters\nbcr{4}{, without performing any preventive refresh}. If a RAT entry is allocated for the row, \X{} increments the RAT counter by one. Otherwise, \X{} increments the row's \nbcr{5}{minimum-valued CT counter by one.}} 

\noindent\nbcr{3}{\head{4. Early \nbcr{6}{Preventive} Refresh \nbcr{6}{at Coarse Granularity} (\secref{sec:earlyref})}
\nbcr{4}{\X{} can perform unnecessary preventive refreshes when a workload hammers many DRAM rows \gls{npr} times \nbcr{5}{such that it causes frequent RAT evictions due to RAT's limited capacity.}
By determining the RAT capacity is insufficient to cover all DRAM rows that \nbcr{5}{reach} \gls{npr} activations in the last \nbcr{5}{counter} reset period, \X{} can refresh DRAM rows and safely reset saturated \nbcr{5}{CT} counters \nbcr{5}{and RAT counters}.
If a DRAM row's CT counters are already at \gls{npr} upon activation, it means that \X{} already identified the DRAM row as an \textit{aggressor row} in earlier cycles of the counter reset period. If there is no RAT entry allocated for such a row, \X{} determines that the row was evicted from RAT due to its limited capacity.}
\X{} keeps track of the number of RAT misses caused by \nbcr{4}{evicted} aggressor rows \nbcr{4}{by flagging any DRAM row that already has \gls{npr} activations in its CT counters upon an activation}. If the number of such \nbcr{4}{misses} exceeds a predetermined threshold, \X{} issues \gls{trefw}/\gls{trefi} refresh commands to refresh \textit{all} DRAM rows in the DRAM rank\footnote{\nbcr{5}{We implement the early \nbcr{6}{preventive} refresh operation with \nbcr{6}{the} rank \nbcr{6}{granularity} $REF$ command. Alternatively, it can be implemented \nbcr{6}{at} bank \nbcr{6}{granularity}. However, DDR4~\cite{jedec2017ddr4} does not support a bank-level $REF$ command. Therefore, a bank-level early \nbcr{6}{preventive} refresh operation requires \nbcr{6}{issuing $ACT$ and $PRE$ commands to all DRAM rows in a bank, which has an additional latency. In future and current chips where bank-level $REF$ command is supported, the overhead of early \nbcr{6}{preventive} refresh can be potentially lower~\cite{chang2014improving}.}}} and reset \textit{all} counters in CT and RAT.}

\head{5. Periodic Counter Reset (\secref{sec:npr})}
\nbcr{4}{It is sufficient for \X{} to track DRAM row activations within the counter reset period by configuring \nbcr{5}{its parameter \gls{npr}, i.e.,} the number of activations targeting a DRAM row that causes a preventive refresh. \X{} safely} resets CT and RAT counters \nbcr{4}{at the end of the reset period. After periodic reset, \X{} \nbcr{5}{returns to the state in} Step 1.}

\setcounter{version}{6}

\subsection{Early \nbcr{6}{Preventive} Refresh \nbcr{6}{at Coarse Granularity}}
\label{sec:earlyref}
\X{} can perform unnecessary preventive refreshes due to \nbcr{4}{its} limited RAT capacity. When the number of \nbcr{4}{rows identified as aggressors (i.e., rows with $Num\_ACT=$\gls{npr}) } exceeds the RAT capacity, aggressor rows can be randomly evicted from RAT. \nbcr{4}{\X{} can determine if a DRAM row is an evicted aggressor row by checking its CT counters and comparing against \gls{npr}: upon an activation, an evicted aggressor row's CT counters are already at \gls{npr}. In contrast, another row's counters can be at most \gls{npr}$-1$ and reach \gls{npr} with the new activation.}

\nbcr{4}{When an evicted aggressor row is activated, \X{} uses CT-based estimation as there is no RAT entry allocated for that row. Since the CT counters already store \gls{npr}, \X{} estimates the activation count as \gls{npr}. Thus, \X{} preventively refreshes the victim rows even though the aggressor row may be activated fewer times than \gls{npr} after its last preventive refresh.}
\nbcr{4}{If \X{} performs such unnecessary preventive refreshes frequently, it degrades system performance. To avoid this, \X{} introduces an early \nbcr{6}{preventive} refresh mechanism to refresh \textit{all} DRAM rows in a DRAM rank and reset \textit{all} CT and RAT counters. Even though the early \nbcr{6}{preventive} refresh mechanism introduces a high latency by refreshing a large number of DRAM rows, it resets the saturated counters and, thus, enables \X{} to avoid repeatedly performing unnecessary refreshes. \X{} utilizes RAT statistics to determine when to perform an early \nbcr{6}{preventive} refresh operation. \X{} classifies RAT misses into two categories: 1) capacity misses (i.e., caused by evicted aggressor rows), and 2) compulsory misses (i.e., caused by new aggressor rows reaching \gls{npr} activations). Capacity misses are identified with a flag indicating the CT counters were already at \gls{npr} before the row activation.}

\nbcr{3}{\X{} allocates a \nbcr{6}{RAT miss} history vector consisting of one bit per RAT miss, indicating either a capacity miss (logic-1) or a compulsory miss (logic-0). \X{} compares the number of capacity misses to a threshold called the \textit{early \nbcr{6}{preventive} refresh threshold} (\textit{EPRT})\nbcr{4}{, which is} determined empirically (\secref{sec:earlyrefresh_sensitivity}). If the number of capacity misses exceeds \textit{EPRT}, \X{} refreshes all DRAM rows in the DRAM rank and resets \textit{all} counters in CT and RAT.}

\setcounter{version}{6}

\subsection{Determining the Preventive Refresh Threshold}
\label{sec:npr}
The preventive refresh threshold (\gls{npr}) should be \nbcr{4}{carefully} selected to prevent \nbcr{5}{a row} from reaching activation count \nbcr{5}{\gls{nrh}} before its victims are refreshed. Since the memory controller performs periodic refreshes to a subset of DRAM rows at a time, DRAM rows are refreshed at different times, and \X{} does not know when each row is refreshed. Therefore, \gls{npr} should be selected such that even without \nbcr{4}{knowing when each row is refreshed}, no row\nbcr{5}{'s activation count} can reach the \gls{nrh}. 

\nbcr{4}{Assume \X{} resets its counters with a reset period of \gls{trefw}. This means between two consecutive periodic refreshes of any DRAM row, \X{} can reset its counters \textit{once}. An \nbcr{5}{attacker can hammer an aggressor row} for \gls{npr}-1 times \nbcr{5}{before \X{} refreshes its victims} and wait for the periodic reset. After \X{} resets its counters, the \nbcr{5}{attacker can hammer the same aggressor row} for \gls{npr}-1 times again,} accumulating $2\times($\gls{npr}$-1)$ activation count \nbcr{5}{before \X{} refreshes} the victim rows. 
To ensure this value does not reach \gls{nrh}, 
\X{} needs to set \gls{npr} to \gls{nrh}$/2$.

\nbcr{4}{Resetting counters more frequently can help reduce counter saturation\nbcr{5}{. H}owever\nbcr{6}{,} it also requires changing the \gls{npr} value. For example, if the reset period is \gls{trefw}$/2$, \X{} can reset its counters \textit{two times} in between two consecutive periodic refreshes of a DRAM row instead of \textit{one time}. Thus, an aggressor row can accumulate $3\times($\gls{npr}$-1)$ activation count without refreshing the victim rows, which requires \gls{npr} to be \gls{nrh}$/3$.}
We adopt the following formula to determine the \gls{npr} value based on the selected $k$ value for a reset period of \gls{trefw}$/k$, similar to prior work~\cite{park2020graphene}.

\begin{equation}
\text{\gls{npr}} = \frac{\text{\gls{nrh}}}{k+1}
\label{eq:npr_equation}
\end{equation}

\section{Security Analysis}
\label{sec:ct-sec}

We analyze the security of \X{} by \nbcr{5}{analyzing} two activation count estimation mechanisms: (1) CT-based estimations and (2) RAT-based estimations.

\head{Security of CT-based estimation}
CT utilizes the Count-Min Sketch (CMS) technique~\cite{cormode2005improved} to \nbcr{4}{efficiently} track activation counts of DRAM rows.
\nbcr{3}{CMS counters are always incremented upon a row activation and \nbcr{4}{are} only reset at periodic counter resets and early periodic refresh operations. CMS can overestimate a DRAM row's activation count when there is a counter collision in its counter group. However, it can \textit{never underestimate} the activation count. Consequently, CMS guarantees \nbcr{4}{that} the estimated activation count of a DRAM row is always \nbcr{4}{greater than or equal to} the \nbcr{5}{row's} \nbcr{4}{actual} activation count. Thus, CT-based estimation ensure that}
no row is activated more than \nbcr{4}{\gls{npr}} times in a \nbcr{4}{counter reset period} \nbcr{4}{\nbcr{5}{before} its victims are refreshed}.

\head{Security of RAT-based estimation}
\nbcr{3}{There are two important operations that modify the RAT counter of an aggressor row. First, a RAT counter is reset (i) when \X{} preventively refreshes the aggressor row's victim rows, (ii) after \X{} performs an early preventive refresh that refreshes all DRAM rows, and (iii) after a periodic reset. \nbcr{4}{In these cases, either the aggressor row's victims are refreshed or \gls{npr} is selected such that the aggressor row cannot be activated \gls{nrh} times without its victims are refreshed.}
Second, a RAT counter is probabilistically evicted when a new aggressor is identified and RAT is full. In this case, the next time the evicted row is activated, \X{} falls back to the CT-based estimation that always \nbcr{5}{provides} an \nbcr{5}{overestimate} of the actual activation count (as \nbcr{5}{described} above).}

\section{Evaluation Methodology}
\label{sec:methodology}

We evaluate \X{}'s performance and energy consumption with Ramulator \cite{kim2016ramulator, ramulatorgithub}, a cycle-level DRAM simulator, and DRAMPower~\cite{drampower}. \nbcr{4}{Table \ref{configs} provides} our simulated system's configuration.

\begin{table}[ht]
\footnotesize
\centering
\caption{Simulated System Configuration}
\label{configs}
\resizebox{\linewidth}{!}{
\begin{tabular}{l|l} 
\hline\hline
\textbf{Processor}        & \begin{tabular}[c]{@{}l@{}}1 or 8 cores, 3.6GHz clock frequency,\\ 4-wide issue, 128-entry instruction window\end{tabular}                                                                                                                      \\ 
\hline
\textbf{DRAM}             & \begin{tabular}[c]{@{}l@{}}DDR4, 1 channel, 2 rank/channel, 4 bank groups,\\ 4 banks/bank group, 128K rows/bank\end{tabular}                                                                                                                    \\ 
\hline
\textbf{Memory Ctrl.}     & \begin{tabular}[c]{@{}l@{}}64-entry read and write requests queues,\\Scheduling policy: FR-FCFS~\cite{rixner2000memory,zuravleff1997controller} \\with a column cap of 16~\cite{mutlu2007stall}
\end{tabular}  \\ 
\hline
\textbf{Last-Level Cache} & 8 MiB (single-core), 16 MiB (8-core)                                                                                                                                                                                                            \\
\hline
\end{tabular}
}
\end{table}

\head{Comparison Points} We compare \X{} to a baseline system with no RowHammer mitigation and to \param{four} state-of-the-art RowHammer mitigation mechanisms.

\textbf{(1)} \textbf{Graphene}~\cite{park2020graphene} employs \nbcr{4}{the} Misra-Gries~\cite{misra1982finding} algorithm to track possible aggressor rows with tagged counter tables per bank. When a counter value exceeds a threshold value, Graphene issues preventive refreshes to the victim rows.

\textbf{(2)} \textbf{Hydra}~\cite{qureshi2022hydra} \nbcr{4}{combines an SRAM-based counter table in the memory controller and a per-DRAM-row counter table in DRAM to incur a low area overhead in the memory controller. First, Hydra implements a \textit{group count table} to track aggregated row activations of row groups with low area overhead. Second, it implements a \textit{row count table} to track per-DRAM-row activations in DRAM. When a row group exceeds a predetermined threshold, Hydra writes the group's aggregated activation count to dedicated per-DRAM-row counters in DRAM and caches per-DRAM-row counters in the memory controller.}  Hydra performs preventive refresh when a \nbcr{4}{dedicated per-DRAM-row} counter exceeds a threshold value. We configure Graphene and Hydra for the tested RowHammer thresholds as described in their original works~\cite{park2020graphene,qureshi2022hydra}.

\textbf{(3)} \textbf{REGA}~\cite{marazzi2023rega} \nbcr{5}{modifies the} DRAM design to concurrently refresh one or more victim rows \nbcr{5}{\emph{every time}} a DRAM row is activated. \nbcr{4}{With decreasing \gls{nrh}, REGA maintains its protection guarantees by refreshing more DRAM rows concurrently with each DRAM activation. To do so, REGA reduces} the default $t_{RC}$ 
value. A smaller $t_{RC}$ 
allows REGA to refresh more rows concurrently with a DRAM row activation, but it also increases the access latency. To simulate REGA, we modify 
$t_{RC}$
as described in~\cite{marazzi2023rega}. 

\textbf{(4)} \textbf{PARA}~\cite{kim2014flipping} prevents \rh{} bitflips by performing preventive refreshes to adjacent rows of a row that is being closed based on a probability threshold. We tune the probability threshold of PARA for a target failure probability of $10^{-15}$ within a \SI{64}{\milli\second} as in prior work~\cite{yaglikci2021blockhammer,yaglikci2022hira}.

\head{Workloads} We evaluate \param{61} single-core and \param{56} \nbcr{4}{homogeneous multi-programmed} 8-core workload from \nb{five} benchmark suites: SPEC CPU2006~\cite{spec2006}, SPEC CPU2017~\cite{spec2017}, TPC~\cite{tpcweb}, MediaBench~\cite{fritts2009media}, and YCSB~\cite{ycsb}. Based on the row buffer misses-per-kilo-instruction (RBMPKI) \nbcr{4}{metric}~\cite{yaglikci2021blockhammer, olgun2024abacus}, we group workloads into three categories, which Table \ref{table:workloads} describes: 1) low memory-intensity, RBMPKI $\in [0,2)$, 2) medium memory-intensity, RBMPKI $\in [2,10)$, 3) high memory-intensity, RBMPKI $\in [10+)$. To do so, we obtain
the RBMPKI values of the applications by analyzing \nbcr{4}{each workload's} SimPoint \cite{calder2005simpoint} traces (200M instructions). \revdt{\revd{2}Table~\ref{table:workloads} shows the average memory bandwidth consumption (in MB/s) of each workload in parentheses.}

\begin{table}[ht]
\caption{Evaluated Workloads \nbcr{4}{and Their Characteristics}}
\label{table:workloads}
\scriptsize
\centering
\resizebox{\linewidth}{!}{
\begin{tabular}{|l||l|}
\hline
\multicolumn{1}{|c||}{\textbf{RBMPKI}}                      & \multicolumn{1}{c|}{\textbf{Workloads}}                                                                                                                                                                                                                                                                             \\ \hline \hline
\begin{tabular}[c]{@{}l@{}}$[10+)$ \\ (High)\end{tabular} & \begin{tabular}[c]{@{}l@{}}

519.lbm \revdt{(5049 MB/s)}, 459.GemsFDTD \revdt{(4788 MB/s)},
 450.soplex \revdt{(3212 MB/s)}, \\ h264\_decode \revdt{(11284 MB/s)},
520.omnetpp \revdt{(2567 MB/s)},
433.milc \revdt{(3595 MB/s)},
\\ 434.zeusmp \revdt{(5115 MB/s)}, bfs\_dblp \revdt{(12135 MB/s)},
429.mcf \revdt{(5588 MB/s)}, \\ 549.fotonik3d \revdt{(4428 MB/s)},
470.lbm \revdt{(6489 MB/s)}, bfs\_ny \revdt{(12146 MB/s)}, 
\\bfs\_cm2003 \revdt{(12138 MB/s)}, 437.leslie3d \revdt{(3806 MB/s)}
\end{tabular}      
\\ \hline
\begin{tabular}[c]{@{}l@{}}$[2, 10)$\\ (Medium)\end{tabular}  & \begin{tabular}[c]{@{}l@{}}
510.parest \revdt{(92 MB/s)}, 462.libquantum \revdt{(6089 MB/s)},
 tpch2 \revdt{(3612 MB/s)}, \\ wc\_8443 \revdt{(1772 MB/s)}, 
ycsb\_aserver \revdt{(1080 MB/s)},
473.astar \revdt{(2473 MB/s)}, 
\\ jp2\_decode \revdt{(1390 MB/s)}, 436.cactusADM \revdt{(1915 MB/s)}, 557.xz \revdt{(1113 MB/s)}, 
\\ycsb\_cserver \revdt{(842 MB/s)}, ycsb\_eserver \revdt{(721 MB/s)}, 
471.omnetpp \revdt{(96 MB/s)}, 
\\483.xalancbmk \revdt{(187 MB/s)},
505.mcf \revdt{(3760 MB/s)}, wc\_map0 \revdt{(1768 MB/s)}, 
\\jp2\_encode \revdt{(1706 MB/s)}, tpch17 \revdt{(2553 MB/s)},
 ycsb\_bserver \revdt{(854 MB/s)}, \\tpcc64 \revdt{(1472 MB/s)},
482.sphinx3 \revdt{(968 MB/s)}
\end{tabular}                                                                                                                      \\ \hline
\begin{tabular}[c]{@{}l@{}}$[0, 2)$\\ (Low)\end{tabular}     & \begin{tabular}[c]{@{}l@{}}
502.gcc \revdt{(180 MB/s)}, 544.nab \revdt{(78 MB/s)},
 h264\_encode \revdt{(0.10 MB/s)},
 \\507.cactuBSSN \revdt{(1325 MB/s)}, 
525.x264 \revdt{(109 MB/s)}, ycsb\_dserver \revdt{(659 MB/s)},
\\531.deepsjeng \revdt{(105 MB/s)}, 526.blender
 \revdt{(56 MB/s)}, 435.gromacs \revdt{(259 MB/s)},
\\523.xalancbmk \revdt{(180 MB/s)}, 
447.dealII \revdt{(24 MB/s)},
508.namd \revdt{(104 MB/s)},
\\538.imagick \revdt{(8 MB/s)},
 445.gobmk \revdt{(97 MB/s)}, 444.namd \revdt{(104 MB/s)}, 
 \\464.h264ref \revdt{(17 MB/s)}, 
ycsb\_abgsave \revdt{(362 MB/s)},
458.sjeng \revdt{(131 MB/s)}, 
\\541.leela \revdt{(4 MB/s)}, 
 tpch6 \revdt{(675 MB/s)}, 511.povray \revdt{(1 MB/s)}, \\
456.hmmer \revdt{(28 MB/s)},  481.wrf \revdt{(7 MB/s)}, 
grep\_map0 \revdt{(381 MB/s)}, \\
500.perlbench \revdt{(642 MB/s)},
403.gcc \revdt{(79 MB/s)}, 401.bzip2 \revdt{(59 MB/s)}

\end{tabular} \\ \hline
\end{tabular}
}
\end{table}

\nbcr{4}{We open-source our simulat\nbcr{5}{ion} infrastructure, \nbcr{5}{workload, scripts} at \url{https://github.com/CMU-SAFARI/CoMeT} to enable reproducibility and aid future research. }
\setcounter{version}{5}
\section{\X{} Exploration \nbcr{4}{and Implementation}}
\label{sec:exploration}

\nbcr{3}{In this section, we 1) explore the design space of \X{}, 2) present \nbcr{4}{its} hardware implementation, and 3) evaluate \X{}'s area overhead}.

\subsection{Sensitivity Analysis}
\nbcr{3}{We explore the design space of \X{} in terms of (i)~the Counter Table (CT) design, (ii)~the Recent \nbcr{4}{Aggressor} Table (RAT) design, (iii)~the early \nbcr{6}{preventive} refresh mechanism design, and (iv)~\nbcr{4}{counter} reset period and \gls{npr} selection.}

\subsubsection{CT Design Space Exploration}
\label{sec:ct-configuration} 
\nbcr{3}{We sweep the number of hash functions ($N_{Hash}$) and the number of counters implemented per hash function ($N_{Counters}$) and show the impact of CT design on \X{}'s performance. }
\figref{fig:ct-sweep} show the \nbcr{4}{normalized} performance distribution of \X{} designs with different ($N_{Hash}$, $N_{Counters}$) pairs for \gls{nrh}$=1K$ (a) and \gls{nrh}$=125$ (b), across 61 single-core applications normalized to a baseline with no \rh{} mitigation. For this experiment, we select a RAT size of 128 to observe the isolated impact of CT design on performance. 

We make \param{three} key observations \nbcr{3}{based on~\figref{fig:ct-sweep}.} 
First, \nbcr{4}{as} $N_{Hash}$ \nbcr{4}{increases,} \nbcr{3}{\X{}'s performance overhead decreases} across all $N_{Counters}$ values \nbcr{3}{at both \gls{nrh} values}. \nbcr{3}{\nbcr{4}{As  $N_{Hash}$ increases,} \X{} maps each DRAM \nbcr{4}{row} to a larger \nbcr{4}{set} of counters. This way, increasing $N_{Hash}$ introduces redundancy in \nbcr{4}{counters with the minimum value} and reduces unnecessary preventive refreshes due to collisions \nbcr{4}{in CT}.} 
Second, \nbcr{4}{as} \nbcr{3}{$N_{Counters}$ \nbcr{4}{increases}, \X{}'s performance overhead decreases across all $N_{Hash}$ values at both \gls{nrh} values. Implementing more counters per hash function distributes DRAM rows across more counters in each row, reducing \nbcr{4}{collisions} in each counter. } 
Third, \nbcr{3}{\nbcr{4}{increasing both $N_{Hash}$ and $N_{Counters}$} does not improve \nbcr{4}{\X{}'s performance beyond} $N_{Hash}=4$ and $N_{Counters}=512$ at \gls{nrh}$=125$. We select this configuration as the \nbcr{4}{one we use in our evaluations}} since it \nbcr{4}{provides the best performance at \nbcr{5}{a low} storage cost}. 

\begin{figure}[h]
        \centering
        \includegraphics[width=\linewidth]{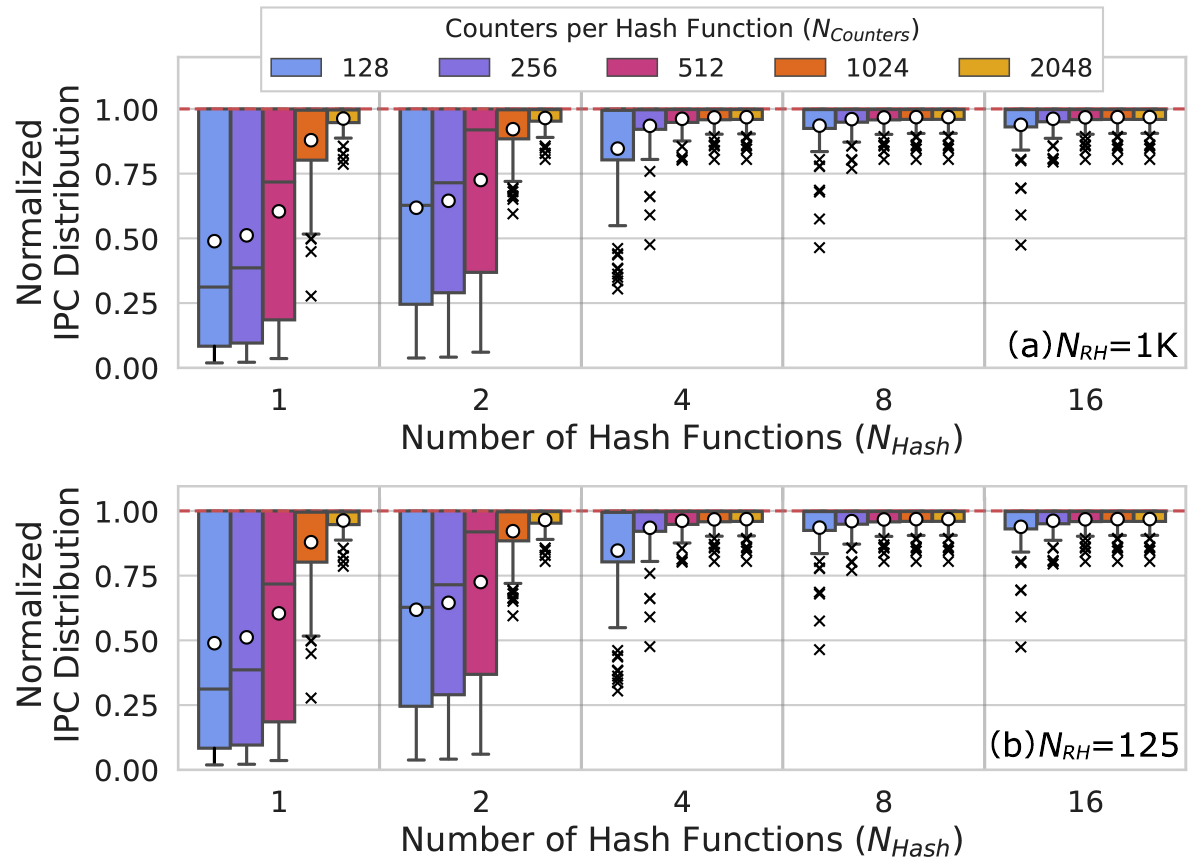}
    \caption{\nbcr{4}{The effect of ($N_{Hash}$, $N_{Counter}$) pairs on \X{}'s performance} at \gls{nrh}$=1K$ (a) and \gls{nrh}$=125$ (b).}
    \label{fig:ct-sweep}
\end{figure}

\subsubsection{RAT Design Space Exploration}
\label{sec:rat-configuration} 
\nbcr{3}{We sweep the number of RAT entries and show the impact of RAT design on \X{}'s performance.} \figref{fig:rat-sweep-125} shows the \nbcr{4}{normalized} performance distribution of \X{} designs with different numbers of RAT entries ($N_{RAT\_Entries}$) for different \gls{nrh} values across 61 single-core applications normalized to a baseline with no \rh{} mitigation. For this experiment, we select a CT design with $N_{Hash}=4$ and $N_{Counters}=512$ to observe the isolated impact of RAT design on performance. 

\begin{figure}[ht]
\centering
\includegraphics[width=\linewidth]{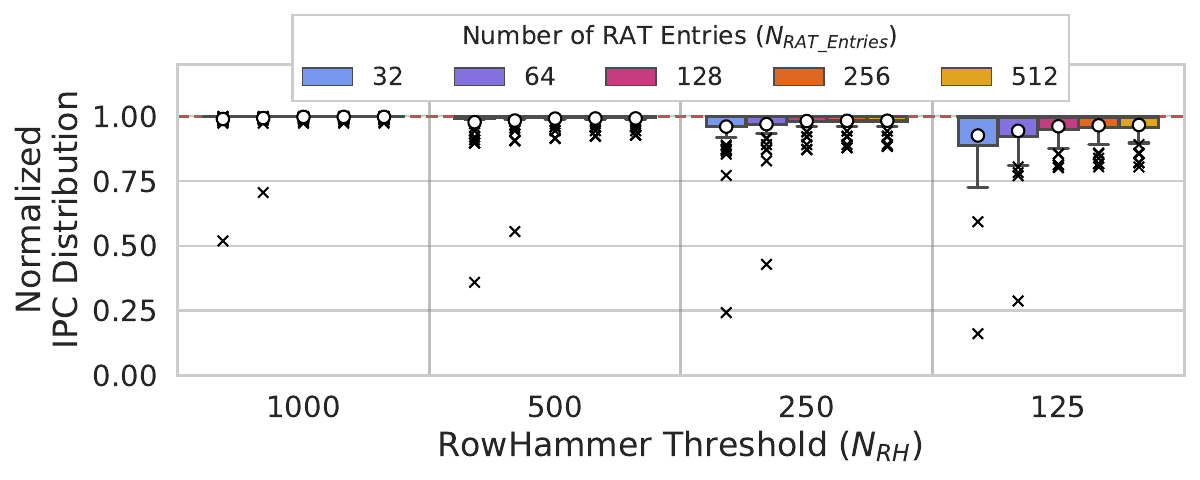}
\caption{\nbcr{4}{The effect of $N_{RAT\_Entries}$ on \X{}'s performance.}}
\label{fig:rat-sweep-125}
\end{figure}

\nbcr{3}{We make two observations based on \figref{fig:rat-sweep-125}. First, increasing the RAT size to more than 128 entries does not significantly improve performance. Therefore, we select \nbcr{5}{a} 128-entry RAT as the \nbcr{4}{one we use in our evaluations}. Second, the \nbcr{4}{effect of the} RAT size on \X{}'s performance increases at low \gls{nrh} values. This is because, \nbcr{4}{at} low \gls{nrh} values, more DRAM rows \nbcr{4}{experience} \gls{npr} activations on average. Thus, \X{} utilizes \nbcr{4}{the} RAT more at low \gls{nrh} values.}

\setcounter{version}{6}

\subsubsection{\nbcr{3}{Early \nbcr{6}{Preventive} Refresh Design Space Exploration}}
\label{sec:earlyrefresh_sensitivity}
\nbcr{3}{We sweep the \nbcr{6}{RAT miss} history length and the \nbcr{6}{early preventive refresh threshold} (\textit{EPRT}) \nbcr{4}{(\secref{sec:earlyref})} and show the impact of the early \nbcr{6}{preventive} refresh mechanism configuration on \X{}'s performance and DRAM energy consumption. \nbcr{4}{We observe that RAT capacity has the most impact on performance for memory-intensive multicore workloads because large numbers of DRAM rows reach \gls{npr} and \nbcr{6}{are} placed in RAT \nbcr{6}{in these workloads}. To stress the RAT capacity, w}e evaluate the performance and DRAM energy consumption of 8-core workloads running on a system with \X{} designs implementing different early \nbcr{6}{preventive} refresh configurations at \gls{nrh}$=125$ \nbcr{4}{(see methodology in~\secref{sec:methodology})}. \figref{fig:early} shows the performance and DRAM energy consumption distribution of 8-core workloads for different RAT miss history lengths and \textit{EPRT} values.}
\nbcr{3}{We set \textit{EPRT} to be proportional to the RAT miss history length such that \X{} performs an early \nbcr{6}{preventive} refresh when the number of RAT capacity misses is 25\%, 50\%, 75\%, and 100\% of all RAT misses in the history vector. 
We evaluate an additional configuration of \X{} that performs an early \nbcr{6}{preventive} refresh whenever a \nbcr{4}{RAT} capacity miss occurs, labeled as 0\%.}

\nbcr{3}{We make four observations. First, with low \textit{EPRT} values (\nbcr{4}{e.g., 0\% \textit{EPRT}}), \X{}'s performance overhead and DRAM energy consumption are high. Performing frequent early \nbcr{6}{preventive} refresh operations is costly because each early \nbcr{6}{preventive} refresh operation (i) stalls many DRAM requests by making the target DRAM rank unavailable and (ii) consumes DRAM energy by refreshing all DRAM rows in all DRAM banks. Second, with high \textit{EPRT} values (such as 75\% and 100\% of all RAT misses), \X{} does not perform many early \nbcr{6}{preventive} refreshes. Thus, workloads that exceed RAT capacity may suffer from performance and energy overheads due to unnecessary preventive refreshes. 
Third, increasing the RAT miss history length improves performance and energy consumption until the history length is 256. Fourth, with a RAT miss history length of 256, \X{} provides the best performance \nbcr{4}{at EPRT$=25\%$} and the least energy cost \nbcr{4}{at EPRT$=50\%$}. \nbcr{4}{This is because \nbcr{6}{as \textit{EPRT} decreases}, \X{} avoids more unnecessary preventive refreshes \nbcr{6}{as it resets} CT and RAT counters earlier. However, at the same time, \X{} can perform \nbcr{6}{more early preventive refresh} operations and induce a higher energy overhead.} We \nbcr{6}{empirically set} the \nbcr{6}{RAT miss history length to} 256 and \nbcr{4}{\textit{EPRT} \nbcr{6}{to} $25\%$ because} \nbcr{6}{these values} provide better performance at the expense of small energy overheads.}

\begin{figure}[h]
        \centering
        \includegraphics[width=\linewidth]{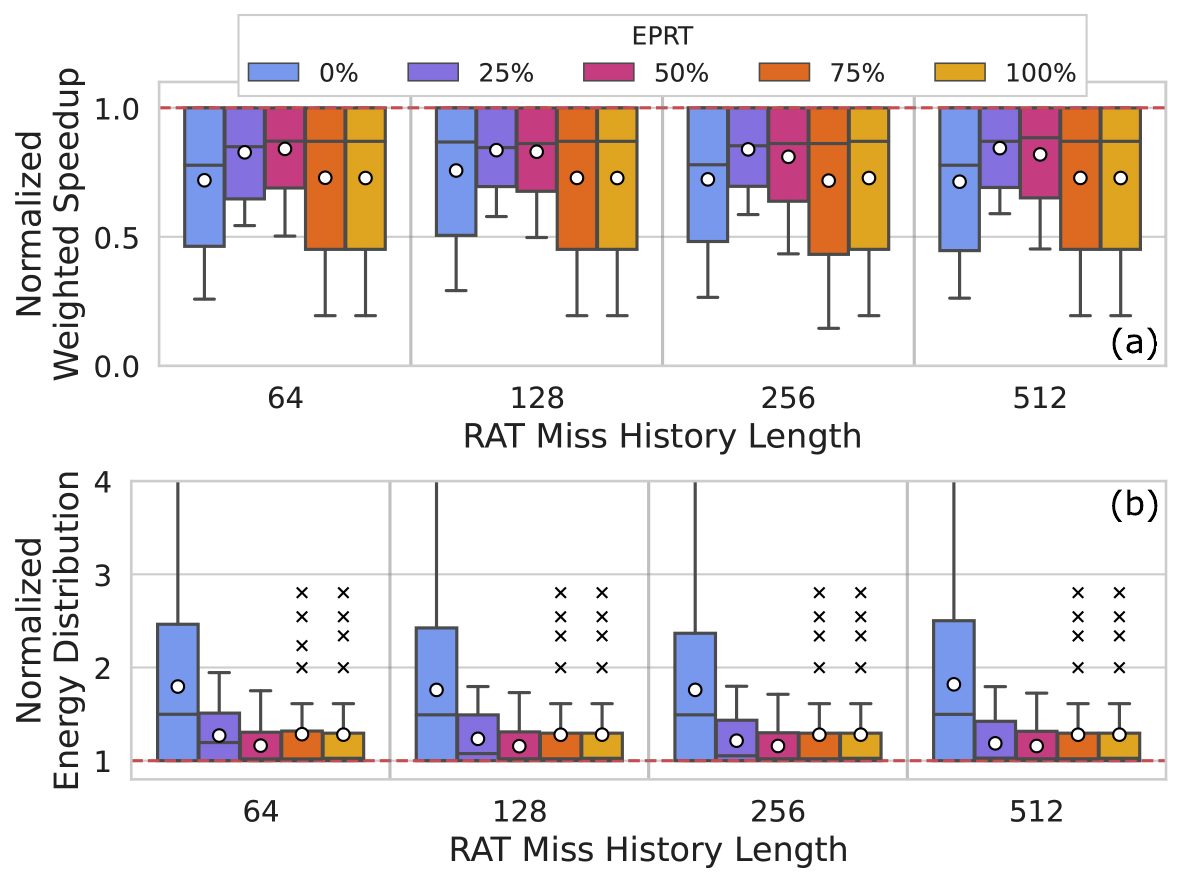}
    \caption{\nbcr{4}{The effect of \emph{EPRT} and RAT Miss History Length} on \X{}'s performance (a) and DRAM energy consumption (b).}
    \label{fig:early}
\end{figure}

\setcounter{version}{6}

\subsubsection{\nbcr{4}{Counter} Reset Period and \gls{npr} Exploration}
\label{sec:npr_eval} 
\nbcr{4}{We set \X{}'s counter reset period to \gls{trefw}$/k$ and \gls{npr} to \gls{nrh}$/(k+1)$, based on Equation~\ref{eq:npr_equation}. W}e sweep $k$ from 1 to 5 to evaluate the impact of the reset period \nbcr{4}{and \gls{npr}} on \X{}'s performance. 
\figref{fig:scaling_n} shows the \nbcr{4}{normalized} performance distribution of single-core benign workloads running in a system with \X{} across different \gls{nrh} values.   
We make two key observations based on \figref{fig:scaling_n}. \nbcr{3}{First, increasing $k$ improves the worst-case performance and reduces the maximum slowdown until $k=3$. This is because \nbcr{4}{with increasing $k$, \X{} resets its counters more frequently and avoids using saturated counters. This improves \X{}'s performance by avoiding unnecessary refreshes when running memory intensive workloads that suffer from counter saturation}.
Second, \nbcr{4}{increasing $k$ beyond 3 degrades \X{}'s performance. For $k>3$, \gls{npr} becomes increasingly small (i.e., as \gls{npr}$=$\gls{nrh}$/(k+1)$). For example, at \gls{nrh}$=125$, $k=4$ results in \X{} performing preventive refreshes whenever a DRAM row's CT or RAT counters reach 25 activations. For $k>3$, \X{} incurs more performance overheads with \emph{necessary} preventive refreshes than the performance improvement of avoiding \emph{unnecessary} ones.} 
We conclude that $k=3$ improves \X{}'s worst-case performance significantly without incurring prohibitive performance overheads on average. Thus, we select $k=3$ \nbcr{4}{in our evaluations}.}

\begin{figure}[ht]
\centering
\includegraphics[width=\linewidth]{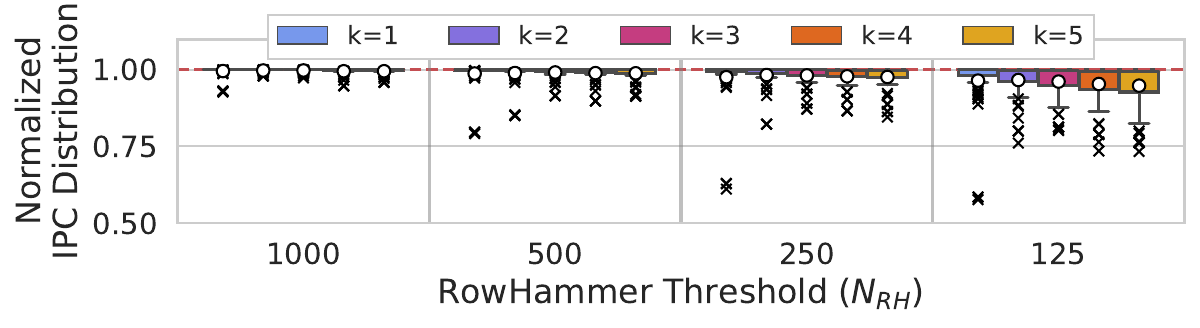}
\caption{Performance overhead with different $k$ values.}
\label{fig:scaling_n}
\end{figure}

\subsection{Hardware Implementation}
\label{sec:hardware}
\X{} is implemented in the memory controller and does not introduce any changes to the DRAM chip \nbcr{4}{or interface}. 

\subsubsection{\X{} Components} 
\X{} employs one Counter Table (CT) and one Recent \nbcr{4}{Aggressor} Table (RAT) per DRAM bank. The CT is structured with 4 rows and 512 columns, implementing 2048 counters in total \nbcr{4}{(\secref{sec:ct-configuration}).} 
\nbcr{3}{Each row of the CT is indexed with different hash functions enabling parallel accesses. Each row is separately implemented using scratchpad SRAM arrays.
RAT consists of 128 entries~\nbcr{4}{(\secref{sec:rat-configuration})}, each comprising a 17-bit RAT tag (row tag bits as specified in \cite{jedec2017ddr4}) and a counter.  We implement RAT using a content-addressable memory (CAM) tag array and a scratchpad SRAM counter array. \X{} allocates 256 bits per DRAM bank for the \nbcr{6}{RAT miss} history vector \nbcr{4}{(\secref{sec:earlyrefresh_sensitivity})}. The RAT miss history vector is implemented as a scratchpad SRAM array.}

\subsubsection{Preventive Refresh Capability} 
\nbcr{3}{The standard refresh ($REF$) command in DRAM standards ~\cite{liu2012raidr,jedec2017ddr4,jedec2020ddr5} is row-address-agnostic. Thus, \X{} cannot use this command to preventively refresh selected DRAM rows. Instead, \X{}}
performs a preventive refresh operation by accessing the target (i.e., victim) row once (i.e., sending $ACT$ and $PRE$ commands), which is compatible with current DRAM specifications. \nbcr{3}{When a DRAM row is identified as an aggressor row, \X{} preventively refreshes the aggressor row's \nbcr{4}{two immediate neighbors (i.e., victims)}. \X{} prioritizes the preventive refreshes over other DRAM requests. This way, the memory controller does not serve any other DRAM request before the victim rows are preventively refreshed.}

\subsection{Area Overhead} 
We evaluate \X{}'s area using CACTI~\cite{cacti} \revdt{and Synopsys Design Compiler~\cite{synopsys}}\revd{1}. \nbcr{4}{We open source our area analysis~\cite{cometgithub}.}
Table~\ref{table:areacost} shows \nbcr{3}{\nbcr{4}{\X{}'s} area and storage cost analysis} and two state-of-the-art counter-based \rh{} mitigation \nbcr{3}{techniques Graphene~\cite{park2020graphene} and Hydra~\cite{qureshi2022hydra} (\secref{sec:methodology}), at different \rh{} thresholds. }

\begin{table}[h!]
\centering
\caption{Dual-rank area overhead of \X{} compared to state-of-the-art \rh{} mitigations.}
\label{table:areacost}
\resizebox{\linewidth}{!}{
\begin{tabular}{|l|r|r|r|r|r|r|r|r|}
\multicolumn{1}{r|}{}                     & \multicolumn{2}{c|}{$N_{RH}$=1K}                   & \multicolumn{2}{c|}{$N_{RH}$=500}                  & \multicolumn{2}{c|}{$N_{RH}$=250}                  & \multicolumn{2}{c|}{$N_{RH}$=125}                   \\ 
\cline{2-9}
\multicolumn{1}{l|}{}                     & \multicolumn{1}{c|}{KB} & \multicolumn{1}{c|}{$mm^2$} & \multicolumn{1}{c|}{KB} & \multicolumn{1}{c|}{$mm^2$} & \multicolumn{1}{c|}{KB} & \multicolumn{1}{c|}{$mm^2$} & \multicolumn{1}{c|}{KB} & \multicolumn{1}{c|}{$mm^2$}  \\ 
\hhline{|=========|}
\textbf{CoMeT}                            & 76.5                    & \revdt{0.09}            & 68.0                    & \revdt{0.08}            & 59.5                    & \revdt{0.07}            & 51.0                    & \revdt{0.07}             \\ 
\hline
CT (SRAM)                                 & 64.0                    & 0.05                     & 56.0                    & 0.05                     & 48.0                    & 0.04                     & 40.0                    & 0.04                      \\ 
\hline
RAT (CAM)                                 & 12.5                    & 0.03                     & 12.0                    & 0.03                     & 11.5                    & 0.03                     & 11.0                    & 0.02                      \\ 
\hline
\revdt{Logic Circuitry}                   & \multicolumn{1}{c|}{-}  & \revdt{0.005}            & \multicolumn{1}{c|}{-}  & \revdt{0.005}            & \multicolumn{1}{c|}{-}  & \revdt{0.005}            & \multicolumn{1}{c|}{-}  & \revdt{0.005}             \\ 
\hhline{|=========|}
\textbf{Graphene~\cite{park2020graphene}} & 207.2                   & 0.49                     & 398.4                   & 1.13                     & 765.0                   & 3.01                     & 1466.2                  & 4.89                      \\ 
\hhline{|=========|}
\textbf{Hydra~\cite{qureshi2022hydra}\footref{footnotehydra}}    & 61.6                    & 0.08                     & 56.5                    & 0.08                     & 51.4                    & 0.07                     & 46.8                    & 0.07                      \\
\hline
\end{tabular}
}
\end{table}

\nbcr{3}{We estimate the area overhead of \X{} to be \revdt{0.09}$mm^2$ per DRAM channel for a dual-rank system at \gls{nrh}$=1K$. 
At \gls{nrh}$=125$, \X{}'s estimated area overhead reduces to \revdt{0.07}$mm^2$. This is because at low \gls{nrh} values, \X{} requires fewer bits for each counter. Since the number of counters is the same across all \gls{nrh} values, \X{}'s area overhead reduces.} \nbcr{4}{We analyze the area overhead of the logic circuitry of \X{}'s components by implementing \X{} in Verilog HDL and synthesizing our design using \nbcr{5}{the} Synopsys Design Compiler~\cite{synopsys} \nbcr{5}{at 65nm}. The logic circuitry of \X{}'s components incurs an area overhead of $<0.005mm^2$.}\footnote{\nbcr{6}{We analyze the area overhead of logic circuitry only for \X{} and ignore this overhead for other mitigation mechanisms in our comparisons.}}

\subsubsection{Area Comparison}
\nbcr{3}{\X{} takes up  $5.4\times$, and $74.2\times$ smaller chip area than Graphene~\cite{park2020graphene} at \gls{nrh}=$1K$ and 125, respectively. Graphene uses tag-based counters implemented as CAMs. With \nbcr{4}{more potential} aggressor rows at lower \gls{nrh}, Graphene has a prohibitively high area cost.}

\nbcr{3}{\X{}'s area overhead is $1.09\times$ that of Hydra's \nbcr{4}{chip area overhead} at \gls{nrh}$=1K$. As \X{}'s area overhead is more sensitive to counter \nbcr{4}{size}, \X{}'s area overhead at \gls{nrh}$=125$ is
\nbcr{4}{$\sim$1\% less than Hydra's} 
\nbcr{4}{while \X{} provides much higher performance than Hydra (see~\secref{sec:results}) with no additional DRAM storage overhead\footref{footnotehydra} at the same time}.}

\nbcr{3}{We compare \X{} with PARA~\cite{kim2014flipping} and REGA~\cite{marazzi2023rega}, which have very low area costs. PARA does not maintain any state. Thus, it has no significant area overhead~\cite{kim2014flipping,yaglikci2021blockhammer}. REGA takes $2.06\%$ DRAM chip area to implement. Compared to these mitigation techniques,} \X{} incurs higher area costs. However, it also incurs significantly lower performance and energy overheads (\secref{sec:results}).

\revdt{\head{Latency Analysis}}\revd{1}
\revdt{We implement \X{} in Verilog HDL and synthesize our design using Synopsys Design Compiler~\cite{synopsys} with a 65 nm process technology to evaluate \X{}'s latency impact on memory accesses. Our evaluation shows that \X{} can be implemented off the critical path in the memory controller. Based on our Verilog design, \X{} estimates a row's activation count in 1.98ns. This latency can be easily overlapped with the latency of regular memory controller operations as it is smaller than $t_{RRD}$ (e.g., 2.5 ns in DDR4~\cite{jedec2017ddr4, micron2014ddr4}).} \nbcr{4}{We open-source our hardware implementation~\cite{cometgithub}.}

\section{Results}
\label{sec:results}

\nbcr{3}{We 1) analyze \X{}'s system performance and DRAM energy overheads and compare \nbcr{4}{them to} the state-of-the-art \rh{} mitigation techniques, 2) analyze \X{}'s performance under \rh{} attacks, and 3) compare \X{}'s \nbcr{4}{CMS-based row activation} tracker to other hash-based trackers.}
\nbcr{4}{We provide more detailed analyses and results in our extended version~\cite{cometarxiv}.}

\subsection{System Performance and DRAM Energy}

\subsubsection{System Performance Overhead}
\figref{fig:singlecore} shows the performance of single-core workloads\footnote{\label{footnote:subset_workload}{Only medium and high RBMPKI workloads (see Table~\ref{table:workloads}) are visible for brevity. GeoMean is across all 61 workloads. \nbcr{4}{We provide \nbcr{5}{our extensive methodology and} results for all workloads in our extended version~\cite{cometarxiv}.}}} for four different near-future and \nbcr{3}{very low} \rh{} thresholds when executed on a system with \X{}, normalized to a baseline system that does not have any \rh{} mitigation. 

We make \param{two} key observations. First, at \gls{nrh}$=1K$, \X{} incurs only a \param{0.19}$\%$ (\param{2.64}\%) average (maximum) performance overhead over the baseline with no \rh{} mitigation. \nbcr{3}{\X{} increases the average memory read latency by  \param{0.18}\% due to preventive refresh operations.}
Second, \X{} prevents bitflips with small performance overhead at very low \rh{} thresholds. \nbcr{4}{At \gls{nrh}$=125$,} \X{} incurs a \param{4.01}$\%$ (\param{19.82}\%) average (maximum) performance overhead over the baseline with no \rh{} mitigation. \nbcr{3}{At \gls{nrh}\nbcr{4}{=125},} \X{} increases the average memory read latency by \param{5.30}\% \nbcr{3}{due to preventive refresh operations. We observe that at very low \gls{nrh}, workloads hammer more DRAM rows (i.e., more DRAM rows reach the preventive refresh threshold) in a reset period. Thus, more CT and RAT counters reach the preventive refresh threshold, and \X{} performs more preventive refreshes. }

\begin{figure}[ht]
\centering
\includegraphics[width=\linewidth]{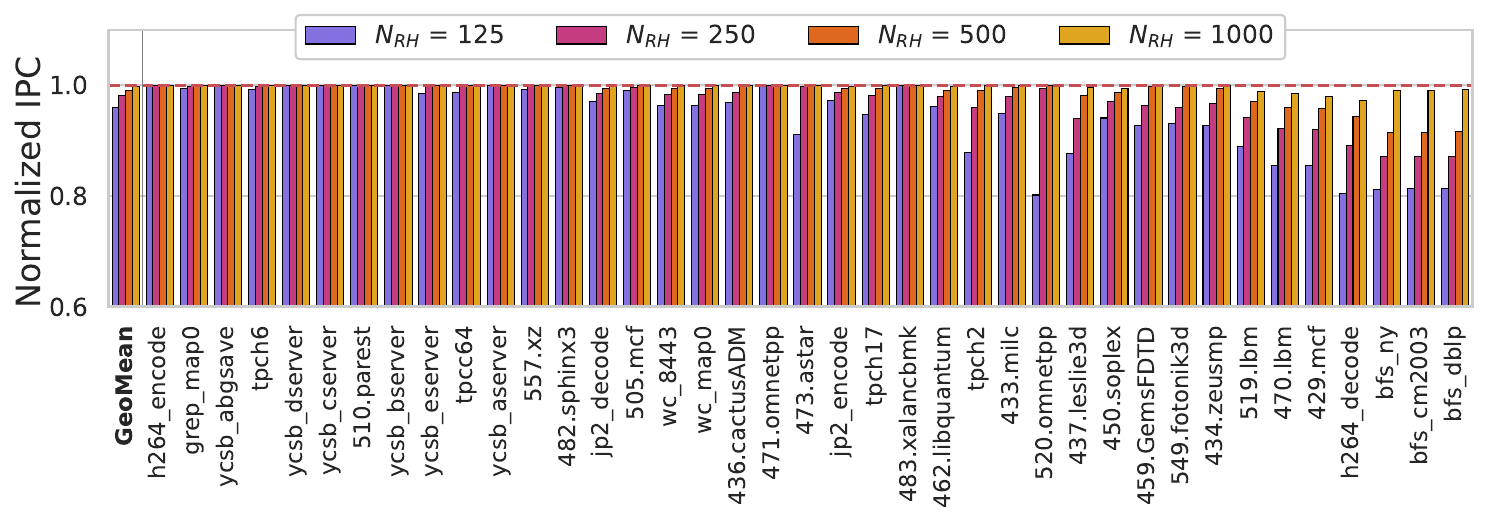}
\caption{\nbcr{4}{Normalized} performance of \nbcr{4}{\X{} on} single-core applications for different \gls{nrh} values.}
\label{fig:singlecore}
\end{figure}

\subsubsection{DRAM Energy Overhead}
\figref{fig:singlecore-energy} shows the DRAM energy consumption for single-core workloads\footref{footnote:subset_workload} for four different \rh{} thresholds when executed on a system with \X{}, normalized to a baseline system that does not have any \rh{} mitigation.

\begin{figure}[ht]
\centering
\includegraphics[width=\linewidth]{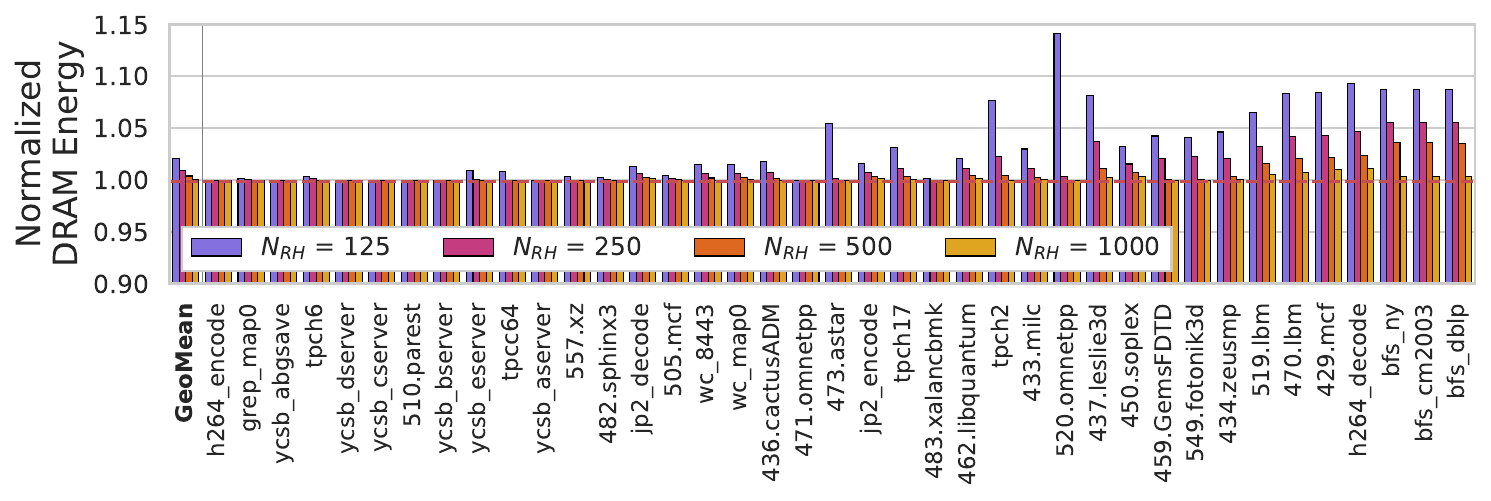}
\caption{\nbcr{4}{Normalized} DRAM energy \nbcr{4}{overhead of \X{} on} single-core applications for different \gls{nrh} values.}
\label{fig:singlecore-energy}
\end{figure}

\nbcr{3}{We make two key observations. First, at \gls{nrh}$=1K$, \X{} increases the DRAM energy consumption by \textit{only} \param{0.08}\% (\param{1.13}\%) on average (at maximum) across all evaluated workloads. Second, at \gls{nrh}$=125$, \X{} increases the DRAM energy consumption by \param{2.07}\% (\param{14.11}\%) on average (at maximum) across all evaluated workloads. This is due to (i) increased DRAM activation and precharge energy induced by the preventive refresh operations and (ii) increased execution time.}

\subsubsection{Performance \nbcr{4}{of \X{} versus State-of-the-Art \rh{} Mitigations}}

\figref{fig:singlecore-comparison} shows the performance comparison of \X{} and four state-of-the-art \rh{} mitigation mechanisms across 61 single-core workloads for four different \gls{nrh} values, normalized to a baseline system without any \rh{} mitigation\nbcr{4}{, as a box plot.\footref{footnote:boxplot}}

\begin{figure}[ht]
\centering
\includegraphics[width=\linewidth]{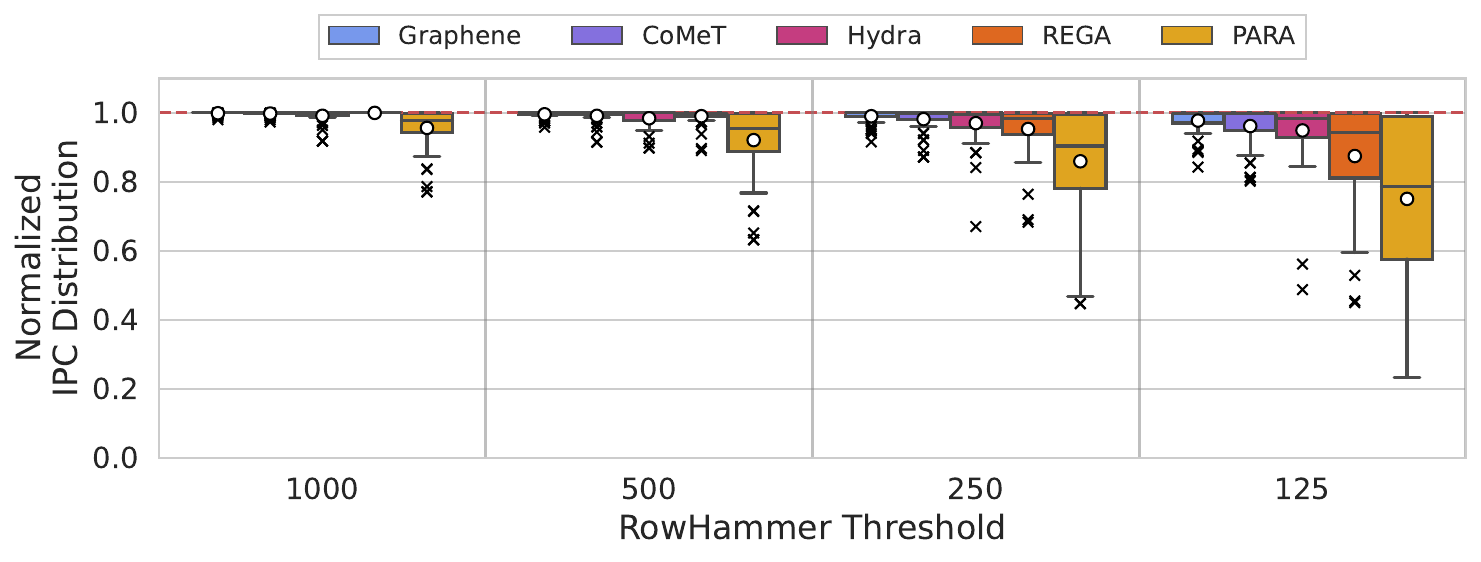}
\caption{Performance comparison of \X{} \nbcr{4}{versus state-of-the-art \rh{} mitigations in single-core workloads.}}
\label{fig:singlecore-comparison}
\end{figure}

We make \param{five} key observations. \nbcr{3}{
First, \X{} outperforms Hydra and PARA at \rh{} thresholds below $1K$ and performs similarly to Graphene at all tested \rh{} thresholds on average across all \nbcr{4}{61} single-core workloads. 
Second, \nbcr{3}{REGA at \gls{nrh}$=1K$ does not incur any performance overhead by performing preventive refreshes concurrently with DRAM row activations. However, REGA incurs increasingly higher performance overheads at lower \gls{nrh} due to the increasing number of preventive refreshes it needs to perform during each DRAM row activation. At \gls{nrh}$=125$, REGA increases $t_{RC}$ significantly and incurs 14.15\% performance overhead on average.} Starting from \gls{nrh}$=500$, \X{} outperforms REGA on average across all workloads.
Third, PARA incurs higher \nbcr{4}{average} performance overheads \nbcr{4}{than} all evaluated \rh{} mitigation techniques due to \nbcr{4}{its very high} preventive refresh probabilities at very low \gls{nrh}. PARA incurs 4.5\% and 30.0\% average performance overhead over the baseline, at \gls{nrh}$=1K$ and \gls{nrh}$=125$, respectively.}
Fourth, 
\X{} demonstrates similar \nbcr{4}{average} performance \nbcr{4}{as} Graphene across all \gls{nrh} values. \nbcr{3}{At \gls{nrh}$=1K$, \X{}'s average performance overhead \nbcr{4}{within} \param{0.08}\% \nbcr{4}{of} Graphene's across all single-core workloads. At \gls{nrh}$=125$, \X{}'s average performance overhead increases to be \nbcr{4}{within} \param{1.75}\% \nbcr{4}{of} Graphene's. We attribute the increase to the unnecessary preventive refresh operations performed by \X{}. \nbcr{5}{Due to hash-based counter collisions,} \X{} overestimates the activation counts of more DRAM rows \nbcr{5}{as \gls{nrh} reduces}. Therefore, at very low \gls{nrh}, \X{} performs more preventive refresh operations and increases the memory read latency compared to Graphene. }
Fifth, \X{} incurs \param{0.67\%} \nbcr{4}{smaller} average performance overhead than Hydra at \gls{nrh}$=1K$. \nbcr{3}{\X{} outperforms Hydra by up to (on average) \param{39.19\%} (\param{1.75\%}) across all workloads at \gls{nrh}$=125$. This is due to the increased off-chip memory requests caused by Hydra. At \gls{nrh}$=125$, Hydra increases the average memory read latency by \param{14.38\%}.}

\figref{fig:multicore-comparison} shows the performance impact of \X{} in terms of weighted speedup~\cite{snavely2000symbiotic,eyerman2008systemlevel,michaud2012demystifying} for four different \rh{} thresholds on an 8-core system, normalized to a baseline system that does not have any \rh{} mitigation.

\begin{figure}[ht]
\centering
\includegraphics[width=\linewidth]{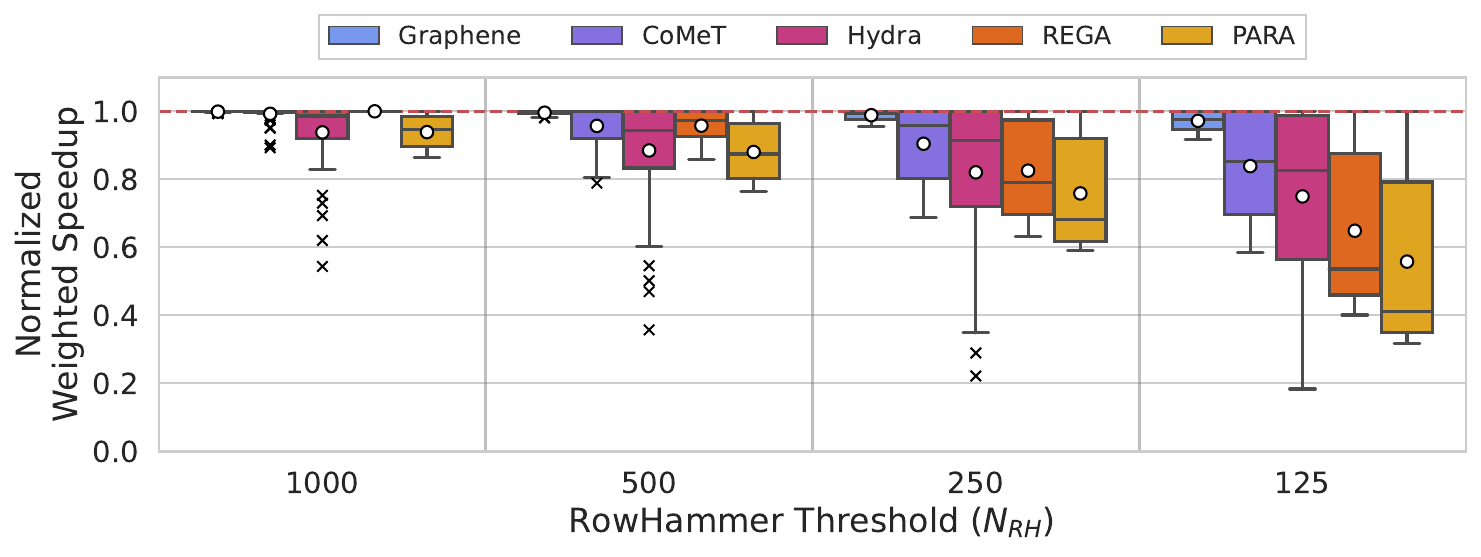}
\caption{Performance comparison of \X{} \nbcr{4}{versus state-of-the-art \rh{} mitigations on} multi-core workloads.}
\label{fig:multicore-comparison}
\end{figure}

We make \param{four} key observations.
First, \X{} induces an average performance overhead of $0.73\%$ at \gls{nrh}$=1K$. \X{}'s average performance overhead increases to $16.11\%$ at \gls{nrh}$=125$.\footnote{\nbcr{4}{We \textit{did not} optimize \X{} for 8-core workloads in our multicore evaluation. Our extended version~\cite{cometarxiv} provides a more detailed multicore analysis.}} At \nbcr{4}{very} low \gls{nrh}, 8-core workloads hammer more DRAM rows more times, causing \X{}'s counters to saturate more quickly. Thus, \X{} performs many \nbcr{3}{unnecessary preventive refreshes. Doing so, at \gls{nrh}$=125$, \X{} increases the average memory request latency by $3.54\times$ over the baseline. }
Second, \X{} outperforms Hydra and PARA at all \rh{} thresholds and REGA starting from \gls{nrh}=250.
Third, \nbcr{3}{\X{} performs similarly to Graphene for \gls{nrh}$=1K$ with an average performance overhead of $0.9\%$ over Graphene. At \gls{nrh}$=125$, \X{}'s average performance overhead is $14.9\%$ higher than Graphene's. This is because \X{} performs \param{46.61\%} more preventive refreshes than Graphene on average across all workloads.
Fourth, at \gls{nrh}$=1K$ \X{} outperforms Hydra by \param{5.82\%} on average across all workloads. At \gls{nrh}$=125$, \X{} outperforms Hydra by up to (on average) \param{$3.19\times$} (\param{11.89\%}). This is due to the large number of off-chip memory requests caused by Hydra. At \gls{nrh}$=125$, Hydra increases the average memory read latency by \param{$5.36\times$} over the baseline due to off-chip memory requests and preventive refreshes.}

\subsubsection{DRAM Energy \nbcr{4}{of \X{} versus State-of-the-Art \rh{} Mitigations}}
\figref{fig:singlecore-energy-comparison} shows the DRAM energy consumption of \X{} and four state-of-the-art \rh{} mitigation techniques on a single-core system for four different \rh{} thresholds, normalized to a baseline system.

\begin{figure}[ht]
\centering
\includegraphics[width=\linewidth]{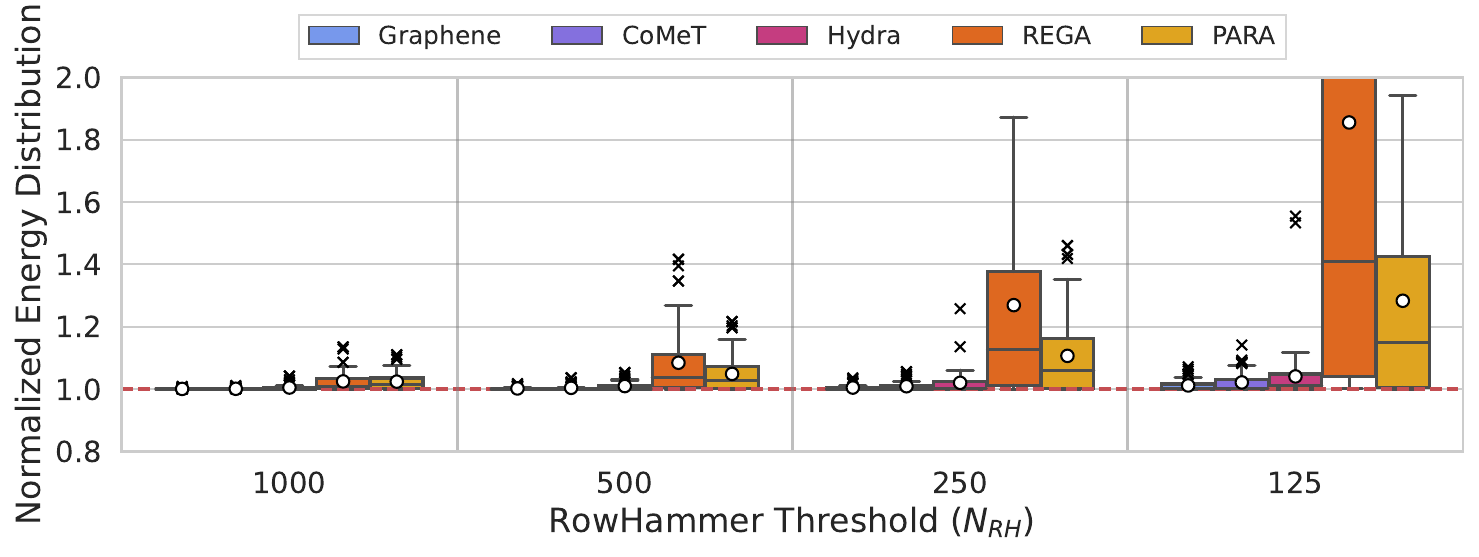}
\caption{DRAM energy \nbcr{4}{overhead} comparison of \X{} \nbcr{4}{versus state-of-the-art \rh{} mitigations on} single-core applications.}
\label{fig:singlecore-energy-comparison}
\end{figure}

We make \param{three} key observations.
\nbcr{3}{First, \X{} consumes less DRAM energy than Hydra, REGA, and PARA on average across all workloads at all \gls{nrh} values.
Second, \X{}'s average DRAM energy overhead is $0.04$\% and $0.96$\% higher than Graphene's across all workloads at \gls{nrh}$=1K$ and 125, respectively.
Third, Hydra incurs \param{0.39}\% and \param{1.88}\% more energy overhead than \X{} on average at \gls{nrh}$=1K$ and \gls{nrh}$=125$, respectively. We attribute this to the increased off-chip memory requests of Hydra.}

\figref{fig:multicore-energy-comparison} shows the DRAM energy consumption of \nbcr{3}{\X{} and four state-of-the-art \rh{} mitigation techniques for four different \rh{} thresholds on an 8-core system, normalized to a baseline system that does not have any \rh{} mitigation. }

\begin{figure}[ht]
\centering
\includegraphics[width=0.95\linewidth]{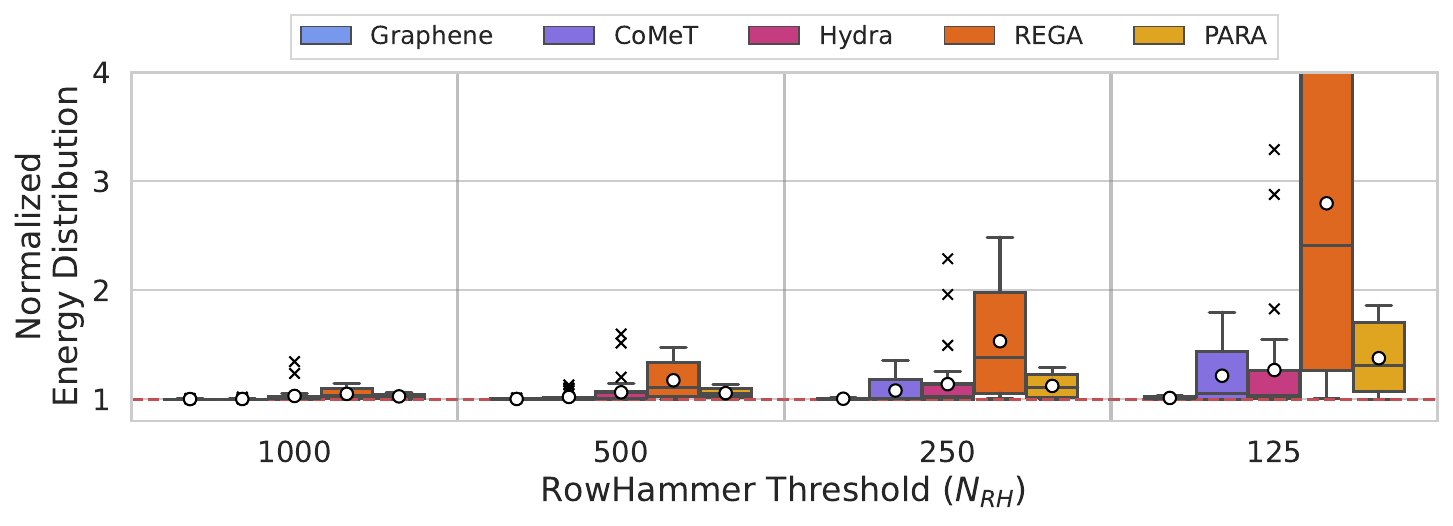}
\caption{DRAM energy \nbcr{4}{overhead} comparison of \X{} \nbcr{4}{versus state-of-the-art \rh{} mitigations on} multi-core workloads.}
\label{fig:multicore-energy-comparison}
\end{figure}

We make \param{four} key observations.
First, \X{} incurs $0.15\%$  more DRAM energy overhead compared to the baseline on average at \gls{nrh}$=1K$. At \gls{nrh}$=125$, \X{}'s average DRAM energy overhead increases to $21.47\%$. \nbcr{3}{This is due to two factors\nbcr{4}{: (1)} at low \gls{nrh} values, 8-core workloads hammer more DRAM rows more times and exceed the RAT capacity by creating a large number of aggressor rows. Therefore, \X{} performs frequent early refresh operations to reset saturated counters, which consumes high DRAM energy. \nbcr{4}{(2)} preventive and early refreshes result in higher execution times, which increases the total DRAM energy consumption.
Second, \X{} incurs \nbcr{4}{lower} average energy overhead \nbcr{4}{than} Hydra, REGA, and PARA at all \rh{} thresholds.
Third, \X{}'s average energy overhead is $0.08\%$ higher than Graphene's for \gls{nrh}$=1K$. At \gls{nrh}$=125$, \X{} incurs $20.87\%$ more average energy overhead than Graphene. This is due to the increased number of preventive and early refresh operations performed by \X{}.
Fourth, Hydra incurs $2.7\%$ and $4.27\%$ more energy overhead than \X{} on average across all workloads at \gls{nrh}$=1K$ and 125, respectively. }

\head{Summary}
\nbcr{3}{We conclude that \X{} prevents \rh{} bitflips at \nbcr{4}{relatively} low performance and energy \nbcr{4}{overheads} on average across all tested single-core and multi-core workloads for \nbcr{4}{most tested} \gls{nrh} \nbcr{4}{values} $1K, 500, 250,$ and 125. \X{} outperforms and consumes less energy than the most-area-efficient state-of-the-art counter-based \rh{} mitigation technique (Hydra~\cite{qureshi2022hydra}). \nbcr{4}{For all tested \rh{} thresholds,} \X{}'s performance and DRAM energy consumption overheads are \nbcr{4}{more similar to} the most-performance-efficient state-of-the-art mitigation technique (Graphene~\cite{park2020graphene}) \nbcr{4}{than other state-of-the-art mitigation techniques (Hydra, REGA, and PARA)}.}

\setcounter{version}{6}

\subsection{Performance Under Adversarial Workloads}\cq{}
\nbcr{3}{We evaluate the performance of \X{} and four state-of-the-art mitigation technique\nbcr{4}{s} for the evaluated single-core workloads that are concurrently running with a RowHammer attack.}
\nbcr{4}{We evaluate} a traditional \rh{} attack \nbcr{6}{running on a single core} that repeatedly activates rows in a refresh period across all banks. \nbcr{4}{We implement this attack in a way that the memory controller issues an ACT command every 20ns while executing the access pattern of the attack trace.}
\figref{fig:attack}\nbcr{4}{(a)} shows the performance overhead of benign applications running concurrently with a \nbcr{3}{traditional} RowHammer attack at \gls{nrh}$=500$. \nbcr{4}{We observe that \X{} incurs a 0.7\% average performance overhead when a \rh{} attack is present and outperforms Hydra, REGA, and PARA, which have \nbcr{6}{average} performance overheads of 2.2\%, 4.0\%, and 14.2\%, respectively.}

\nbcr{4}{\X{} and Hydra have filtering structures (i.e., \X{}'s RAT and Hydra's group counter table) that reduce the performance overhead of their corresponding mechanisms. An attacker can target these structures to cause the \rh{} mitigation mechanism to induce high performance overheads by incurring either (1) a large number of preventive refreshes or (2) a large number of main memory accesses. We implement two targeted attack patterns for \X{} and Hydra that aim to stress their filtering mechanisms and induce high performance overheads. For \X{}, we implement an attack that incurs many RAT evictions and triggers many early preventive refreshes. For Hydra, we implement an attack that incurs excessive off-chip communication by saturating Hydra's group counters.~\figref{fig:attack}(b) shows the performance overhead of \X{} and Hydra for benign applications running concurrently with their corresponding targeted attacks. We observe that \X{} outperforms Hydra by 42.1\%, on average, across all workloads. } 

\begin{figure}[ht]
\centering
\includegraphics[width=0.9\linewidth]{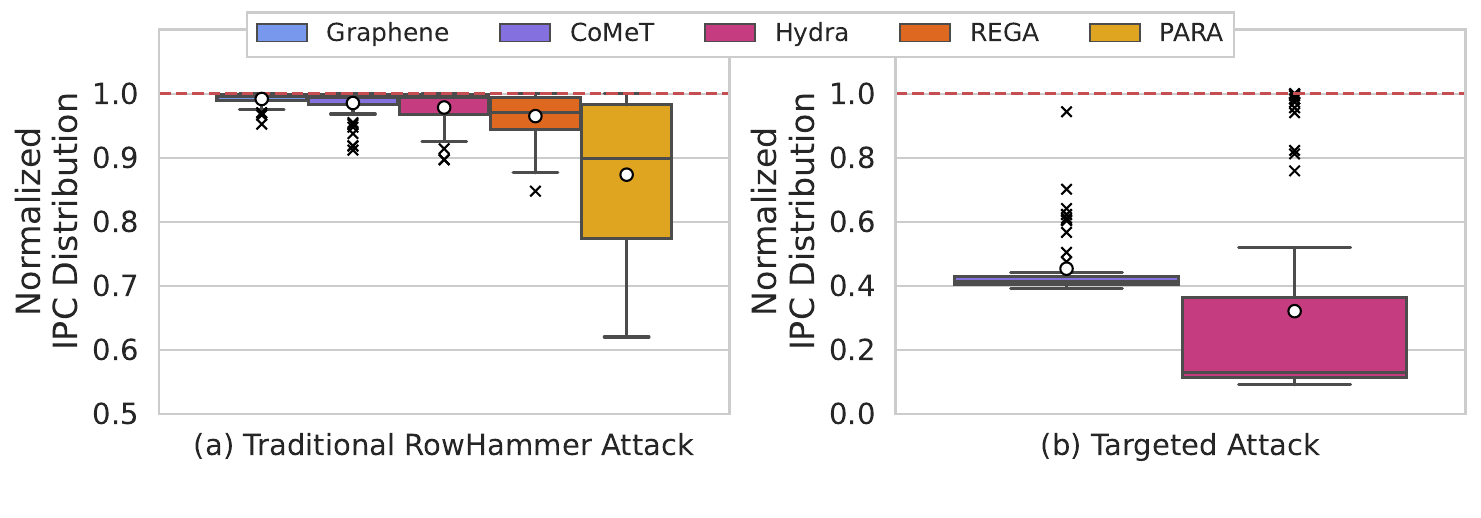}
\caption{\nbcr{4}{Normalized performance of \X{} versus state-of-the-art \rh{} mitigations on (a) a traditional \rh{} attack and (b) a targeted attack.}}
\label{fig:attack}
\end{figure}

\nbcr{3}{We conclude that \X{} incurs negligible additional performance overhead on benign workloads when a traditional \rh{} attack is running at the same time. Targeted attacks can degrade system performance by triggering frequent preventive and early preventive refreshes.}

\setcounter{version}{5}

\subsection{Comparison Against BlockHammer}
\revet{BlockHammer~\cite{yaglikci2021blockhammer} is a \nbcr{4}{\rh{}} mitigation that throttles aggressors. To do so, BlockHammer tracks DRAM rows' activation rates using counting Bloom filters (CBFs)~\cite{bloom1970space, fan2000summary}. CBFs map DRAM rows to different counters by using hash functions. The main difference between \X{} and BlockHammer's trackers is their algorithms, which result in different counter-to-row mapping\nbcr{4}{s} and higher false positive rates for BlockHammer.  BlockHammer's hash functions can map a row to \emph{any} counter in the counter array. In contrast, \X{}'s counter table is divided into different sets, and each hash function can \emph{only} map counters in \emph{its own set} to a row.}

\revet{\figref{fig:blockhammer_fp} shows the false positive rate of BlockHammer's and \X{}'s trackers. The x-axis shows the number of unique rows accessed within one refresh period, from 10 to 100,000. The red vertical line (labeled as $>1$) shows the average number of unique rows touched \textit{at least once} by benign workloads, and the orange vertical line (labeled as $>125$) shows the number of unique rows that are activated at least 125 times. A point in the figure shows a tracker's false positive rate when a total of 10,000 activations\footnote{\nbcr{4}{We observe the \nbcr{5}{total} number of accesses in a refresh window is 10,000 \nbcr{5}{on average across} benign single-core workloads we evaluate}.} are distributed across a number of unique rows. We observe that \X{}'s false positive rate is 41.8\% (21.9\%) smaller than BlockHammer's false positive rate when tested with 100 (250) unique rows. We conclude that when tracking at most \nbcr{4}{2,500 (shown with blue dashed line)} unique rows, \X{}'s tracker outperforms BlockHammer, and they have similar false positive rates when more unique rows are tracked.}
\begin{figure}[ht]
\centering
\includegraphics[width=0.95\linewidth]{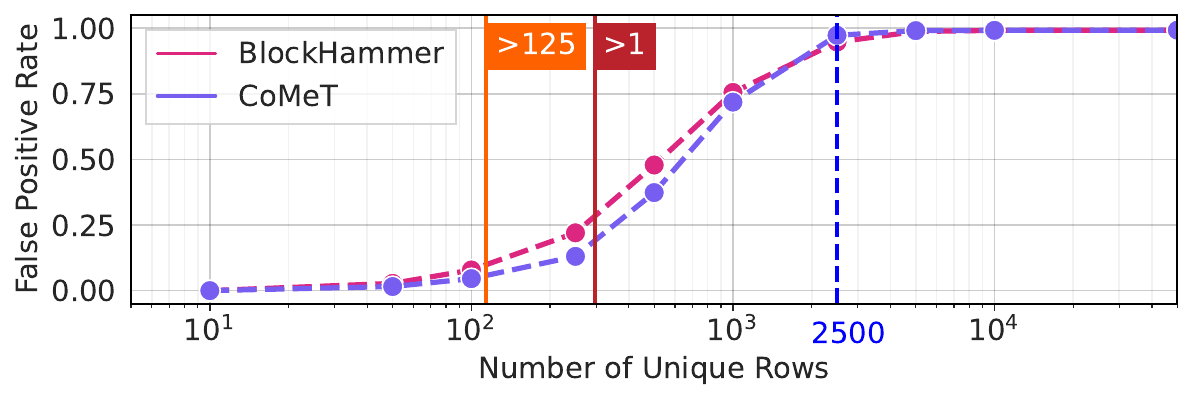}
\caption{\nbcr{4}{\X{} versus BlockHammer} positive rate comparison}
\label{fig:blockhammer_fp}
\end{figure}

\nbcr{3}{\figref{fig:blockhammer_perf} shows the performance comparison of \X{} against BlockHammer on a single-core system across 61 single-core workloads normalized to the baseline design. }We observe that \X{} outperforms BlockHammer by 9.5\% on average, at \gls{nrh}$=125$. This is due to (1) the high false positive rate of BlockHammer and (2) increased memory request latencies caused by throttling. We conclude that BlockHammer has a significant performance overhead at low \rh{} thresholds compared to \X{}.

\begin{figure}[ht]
\centering
\includegraphics[width=0.9\linewidth]{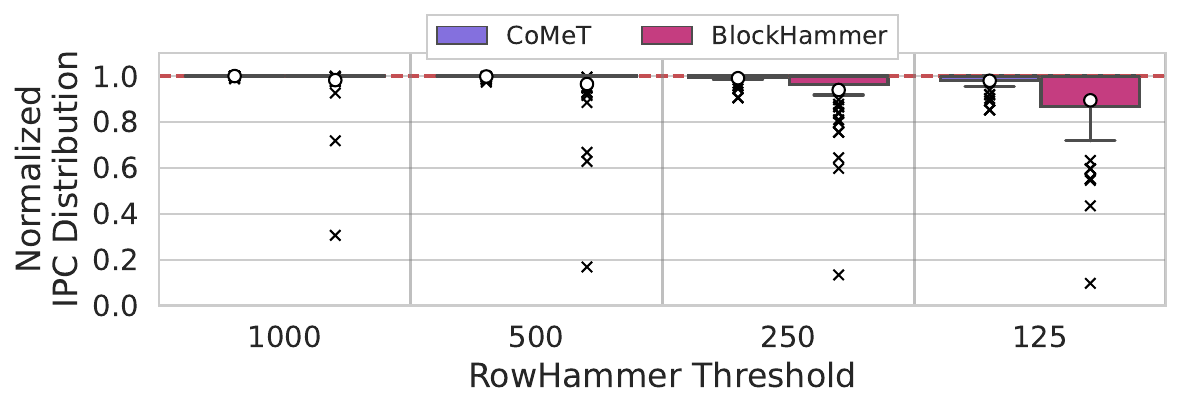}
\caption{Performance comparison against BlockHammer~\cite{yaglikci2021blockhammer}.}
\label{fig:blockhammer_perf}
\end{figure}

\revet{\subsection{\X{} at High \rh{} Thresholds}
We evaluate the performance of 61 single-core applications at high \rh{} thresholds of 2000 and 4000. We observe that \X{} incurs $0.015$\% and $0.0053$\% average performance overheads, at \gls{nrh}$=2000$ and \gls{nrh}$=4000$, respectively. We conclude that \X{} has negligible performance overhead at high \rh{} thresholds.}

\section{Related Work}
\nbcr{4}{To our knowledge, this is the first work to track DRAM row activations with a \nbcr{5}{Count-Min-Sketch}-based technique to prevent \rh{} bitflips with low area, performance, and energy overheads at very low \rh{} thresholds (e.g., 125).  We already} qualitatively and quantitatively compare \X{} with the most relevant \nbcr{6}{state-of-the-art} mechanisms~\cite{park2020graphene,qureshi2022hydra,marazzi2023rega,kim2014flipping} \nbcr{4}{in \secref{sec:results}}. This section discusses other \rh{} mitigation \nbcr{6}{and detection} mechanisms.

\revdt{\head{On-die Mitigation Techniques}\revd{3}
DRAM manufacturers implement RowHammer mitigation techniques, also known as Target Row Refresh (TRR), in commercial DRAM chips~\cite{jedec2020ddr5,jedec2017ddr4}. \nbcr{4}{T}he specific designs of these techniques are not openly disclosed. Recent research shows that custom attacks can bypass these mechanisms~\cite{frigo2020trrespass, hassan2021utrr, jattke2022blacksmith, deridder2021smash, van2016drammer, saroiu2022price} and cause \rh{} bitflips.
In contrast to TRR, \X{} is completely secure (\secref{sec:ct-sec}). Therefore, adopting \X{} is sufficient to securely prevent \rh{} bitflips \nbcr{4}{at} low cost and \textit{without} on-die mitigations. 
To adopt \X{} in a system with TRR, \X{} needs to track target row refreshes. Similar to \nbcr{4}{other mitigation proposals} implemented in the memory controller, \X{} requires information on how TRR works and which rows it refreshes to accurately track activation counts. This information can be provided to \X{} to securely mitigate \rh{}.
}

\head{Hardware-based Mitigation Techniques}
Prior works propose a set of hardware-based mitigation techniques~\citeHardWareBasedMitigations{} to prevent \rh{} bitflips. A subset of these works~\cite{kim2014flipping, you2019mrloc, son2017making, wang2021discreet, yaglikci2022hira, saroiu2022configure,kim2021hammerfilter} proposes probabilistic preventive refresh mechanisms to mitigate \rh{} at low area cost. However, these mechanisms do not provide deterministic \rh{} prevention, in contrast to \X{}. Similarly, three prior works~\cite{joardar2022learning,joardar2022machine,naseredini2022alarm} propose machine-learning-based mechanisms that are not fully secure for all \gls{nrh} values. In contrast, \X{} provides deterministic \rh{} prevention with low area, performance, and energy costs. \nbcr{4}{Three} prior works~\cite{kim2015architectural, qureshi2022hydra,kim2014flipping} propose implementing per-row counters to accurately track row activations. These works incur increasingly large metadata overhead and either cause large area or performance and energy overheads as we show in \secref{sec:motivation}. 
Another group of prior works~\cite{seyedzadeh2017cbt, seyedzadeh2018cbt, kang2020cattwo, lee2019twice, saileshwar2022randomized, saxena2022aqua, kim2022mithril, marazzi2022protrr, park2020graphene, woo2022scalable,olgun2024abacus} propose using \nbcr{4}{the Misra-Gries} frequent item counting algorithm ~\cite{misra1982finding}, which require\nbcr{4}{s a large number of} counters implemented with CAMs for low \rh{} thresholds (\secref{sec:motivation}).
Another set of prior works~\cite{yaglikci2021blockhammer, kim2014flipping, greenfield2012throttling} proposes throttling mechanisms that delay memory requests to prevent \rh{} bitflips. Unlike these mechanisms, \X{} does not throttle any memory requests as it is \nbcr{4}{preventive} refresh-based.
Several works~\cite{bennett2021panopticon, yang2016suppression, hassan2019crow, gomez2016dummy, han2021surround, ryu2017overcoming, zhou2022ltpim, lee2021cryoguard, hassan2022case, devaux2021method, hong2023dsac,kim2023isscc} propose changes in DRAM chip to prevent \rh{} bitflips. These works introduce intrusive changes to the DRAM design and cannot be applied to existing DRAM chips.

\nbcr{3}{A concurrent work, ABACuS~\cite{olgun2024abacus}, proposes a Misra-Gries-based \rh{} mitigation that leverages benign workloads' memory access patterns and modern memory address mapping schemes to reduce the area overhead at the system level. ABACuS uses a single shared activation counter to track DRAM rows that share a row ID across all DRAM banks. By doing so, it incurs a lower area cost in total compared to Graphene~\cite{park2020graphene} and improves scalability \nbcr{4}{with higher} numbers of banks. In contrast, \X{} reduces the per-bank area overhead significantly and improves scalability at low \rh{} thresholds. \X{} and ABACuS are \nbcr{4}{complementary}, and they can be combined to reduce the area overhead at the system level and at low \rh{} thresholds. }

\head{Software-based Mitigation Techniques}
{Several software-based \rh{} mitigation techniques~\cite{konoth2018zebram, van2018guardion, brasser2017can, bock2019riprh, aweke2016anvil, zhang2022softtrr, enomoto2022efficient} propose to avoid hardware-level modifications. However, \nbcr{4}{these works cannot monitor \textit{all} memory requests and thus, \nbcr{5}{many} of them are shown to be defeated by recent attacks~\cite{qiao2016new, gruss2016rowhammer, gruss2018another, cojocar2019eccploit, zhang2019telehammer, kwong2020rambleed, zhang2020pthammer}.}

\head{{Integrity-based Mitigation Techniques}}
Another set of mitigation techniques~\citeIntegrityBasedMitigations{} implements integrity check mechanisms that identify and correct potential bitflips. However, it is either impossible or too costly to address all potential \nbcr{4}{\rh{}} bitflips through these mechanisms. 

\head{RowHammer Detection Mechanisms}
\nbcr{3}{A prior work~\cite{arikan2022processor} proposes a \rh{} detection mechanism with a security checker module based on Count-Min Sketch~(CMS) in embedded systems. It {only} {detects} \rh{} based on predefined instruction patterns by monitoring fetched instructions. In contrast, \X{} mitigates \rh{} by issuing preventive refreshes using its CMS-based DRAM row activation tracker.}

\section{Conclusion}
\nb{In highly \rh{}-vulnerable DRAM-based systems, existing \rh{} mitigations either incur high area overheads or degrade performance significantly. }We propose a new \rh{} mitigation mechanism, \X{}, that prevents \rh{} bitflips with low area and performance cost in DRAM-based systems at \nbcr{4}{very} low \rh{} thresholds \nbcr{4}{(e.g., when 125 activations to the same row can cause a bitflip)}.
The key idea of \X{} is to use low-cost and scalable hash-based counters to track DRAM rows' activation counts by employing \nb{the Count-Min Sketch \nbcr{4}{technique} and reduce the need for expensive fine-grained \nbcr{4}{per-DRAM-}row tracking. \X{} tracks DRAM rows in hash-based counters and allocates per-\nbcr{4}{DRAM-}row counters for only a small fraction of DRAM rows. Thus, \X{} prevent\nbcr{4}{s} \rh{} bitflips at low area, performance, and energy cost\nbcr{5}{s}\nbcr{4}{, as our results demonstrate.}
}

\section*{{Acknowledgments}}
{We thank the anonymous reviewers of HPCA 2024 for their \nbcr{4}{constructive} feedback. 
We thank the SAFARI Research Group members for providing a stimulating intellectual environment. We acknowledge the generous gifts from our industrial partners\nbcr{4}{; especially} Google, Huawei, Intel, \nbcr{4}{and} Microsoft. This work is supported in part by the Semiconductor Research Corporation, the ETH Future Computing Laboratory, \nbcr{4}{Google Security and Privacy Research Award, and the Microsoft Swiss
Joint Research Center}.}

\balance
\bibliographystyle{IEEEtran}
\bibliography{ref}

\clearpage
\appendix

\nobalance

\section{{Artifact Appendix}}

\subsection{Abstract}

{Our artifact provides the source code and necessary instructions to reproduce its key performance and DRAM energy results. We provide: 1) the source code of \X{} implemented using Ramulator~\cite{kim2016ramulator, ramulatorgithub}, 2) scripts to obtain the memory traces of evaluated applications, and 3) scripts to reproduce all key figures in the paper.} We identify the following as key results:
{
\begin{itemize}
    \item Single-core performance evaluation of \X{}
    \item Single-core performance comparison of \X{} against four state-of-the-art \rh{} mitigations: Graphene~\cite{park2020graphene}, Hydra~\cite{qureshi2022hydra}, REGA~\cite{marazzi2023rega}, and PARA~\cite{kim2014flipping}.
    \item  Single-core DRAM energy evaluation of \X{}
    \item Single-core DRAM energy consumption comparison of \X{} against four state-of-the-art \rh{} mitigations.
    \item Sensitivity Analysis of \X{} (Design space exploration of Counter Table and Recent Agressors Table).
    \item Performance evaluation of \X{} across different \gls{npr} and reset window values.
\end{itemize}}

\subsection{Artifact Checklist (Meta-information)}

{\small
\begin{itemize}
  \item {\bf Program:} C++ program, Python scripts (optional Jupyter notebooks), shell scripts.
  \item {\bf Compilation:} Docker-based installation, or G++ version above 8.4. 
  \item {\bf Run-time environment: } Docker-based environment, or Ubuntu 20.04 (or similar) Linux with Python3.9+ and Slurm 20+.
  \item {\bf Execution: } Slurm-based execution.
  \item {\bf Metrics: } Normalized IPC, Normalized DRAM energy consumption.
  \item {\bf Output: } 9 figures in PDF format and related data in plaintext and CSV files.
  \item {\bf How much disk space required (approximately)?: } 20GB
  \item {\bf How much time is needed to prepare workflow (approximately)?: } $\sim1$ hour, depending on the time to download the traces.
  \item {\bf How much time is needed to complete experiments (approximately)?:}  1-24 hours per experiment (depending on the simulated design and workload), and 6405 experiments in total.
  Total completion time: 1-2 days.
  \item {\bf Publicly available?: } Yes
  \item {\bf Archived (provide DOI)?: } \url{10.5281/zenodo.10120298}
\end{itemize}
}

\subsection{Description}
\subsubsection{How to Access} The source code can be downloaded either from GitHub (\url{https://github.com/CMU-SAFARI/CoMeT}) or Zenodo (\url{https://zenodo.org/records/10120298}).

\subsubsection{Hardware dependencies}
To enable easy reproduction of our results, we provide SSH access to our internal Slurm-based infrastructure with all the required hardware and software during the artifact evaluation. Please contact us through HotCRP and/or the AE committee for details.

If the reader will be running the experiments in their own systems or compute clusters:
\begin{itemize}
\item We will be using Docker images to execute experiments. These Docker images assume x86-64 systems.
\item The experiments have been executed using a Slurm-based infrastructure. We \textbf{strongly} suggest using such an infrastructure for bulk experimentation due to the number of experiments required. However, our artifact provides scripts for 1) Slurm-based and 2) native execution.
\item Each experiment takes 1 hour to 24 hours, depending on the simulated design and workload. 
\item CPU traces require $\sim14$GB of storage space.
\end{itemize}

\subsubsection{Software dependencies}
To execute experiments with Docker, we need the following software:
\small{
\begin{itemize}
    \item docker \{docker-ce, docker-ce-cli,containerd.io\} (Tested with Docker version 20.10.23, build 7155243)
    \item curl (Tested with curl 7.81.0)
    \item tar (Tested with tar (GNU tar) 1.34)
\end{itemize}}

We use Docker images that will be downloaded automatically by provided scripts to satisfy the following dependencies:
\small{
\begin{itemize}
    \item GNU Make, CMake 3.10+
    \item G++ version above 8.4
    \item Python 3.9 with Jupyter Notebook
    \item pip packages: pandas, seaborn, and matplotlib
\end{itemize}}

\subsubsection{Data sets} Our docker set-up scripts automatically download and place the traces under \texttt{cputraces/}. Alternatively, the CPU traces can be obtained using \texttt{./get\_cputraces.sh} command. This command fetches the compressed CPU traces and extracts them to \texttt{cputraces/} directory. 

\subsection{Installation}

No system-level installation is required if the reader is accessing our infrastructure for the evaluation or using Docker-based execution. For readers who wish to replicate our results on their own systems without Docker, please 1) follow the instructions in \texttt{README.md} to install all dependencies to run Ramulator, and 2) install Python and pip packages given in the software dependencies. (The reader can use the \textbf{build.sh} script to install all dependencies.)

\subsection{Experiment Workflow}

Our artifact contains 1) the modified source code of Ramulator that implements \X{} and state-of-the-art \rh{} mitigations, 2) scripts to run experiments and collect statistics for native and Slurm-based infrastructures, and 3) python scripts and Jupyter notebooks to plot all figures.

\subsubsection{Launching Experiments}
The following instructions are for launching experiments using the provided scripts for Docker-based execution. 

\head{Launch Experiments in Slurm}

\fancycommand{\$ ./run\_artifact.sh --slurm docker}

\noindent
This command will 1) fetch the Docker image, 2) compile Ramulator inside Docker, 3) fetch CPU traces, and 4) queue Slurm jobs for experiments. 

We suggest using \texttt{tmux} or similar tools that enable persistent bash sessions when submitting jobs to Slurm to avoid any interruptions during the execution of this script.

\head{Launch Experiments in Native Execution}

\fancycommand{\$ ./run\_artifact.sh --native docker}

\noindent
This command will 1) fetch the Docker image, 2) compile Ramulator inside Docker, 3) fetch CPU traces, and 4) run {all} experiments.

Given that this script will run all experiments simultaneously, the reader can modify genrunsp\_docker.py to comment out some configurations to run a subset of experiments at once.

\head{Experiment Completion}

\noindent We expect each experiment to be completed within 24 hours at the latest. Executing all jobs can take 1.5-2 days in a compute cluster, depending on the cluster load. The reader can check the results and statistics generated by the experiments by checking the \textbf{ae-results/} directory. Each experiment generates a file that contains its statistics (\textbf{ae-results/$<$config$>$/$<$workload$>$/DDR4stats.stats}) when it is completed.  

\subsubsection{Obtaining figures and key results}
\head{Docker}
To plot all figures at once using a docker image, the reader can use the \textbf{plot\_docker.sh} script.

\fancycommand{./plot\_docker.sh docker}

This script pulls a Docker image with the Python dependencies. It then plots all figures and saves the results mentioned in the paper under \textbf{plots/} directory.

This command creates the following plots and their related results that are mentioned in the paper:
\begin{itemize}
\item \textbf{comet-singlecore.pdf:} Figure 10
\item \textbf{comet-singlecore-energy.pdf:} Figure 11
\item \textbf{comet-singlecore-comparison.pdf:} Figure 12
\item \textbf{comet-singlecore-energy-comparison.pdf:} Figure 13
\item \textbf{comet-k-evaluation.pdf:} Figure 9
\item \textbf{comet-motiv.pdf:} Figure 3
\item \textbf{comet-ctsweep-1k.pdf:} Figure 6a
\item \textbf{comet-ctsweep-125.pdf:} Figure 6b
\item \textbf{comet-ratsweep.pdf:} Figure 7
\end{itemize}

\head{Non-Docker}
If the reader is not using docker, they can also plot all figures and obtain the results with the following commands:

\fancycommand{\$ cd scripts/artifact/fast-forward/}

\fancycommand{\$ python3 -W ignore create\_all\_results.py }

The reader can specify a result directory with command line argument  \texttt{-r results\_dir}. By default, the script looks for statistics under \textbf{ae-results/}.

\subsection{Evaluation \& Expected Results}
Running the experiments and \textbf{create\_all\_plots.py} is sufficient to regenerate Figures 3,6,7,9-13 and the represented results. The Python script provides further directions to examine the numbers in the paper for each experiment. The reader can check the following results:

\begin{itemize}
    \item \textbf{Single-core Performance Evaluation:} \X{} incurs a 0.19\% (2.64\%) and  4.01\% (19.82\%) average (maximum) performance overhead  at \gls{nrh}=1K an \gls{nrh}=125, respectively, compared to the baseline with no \rh{} mitigation. (plots/singlecore-performance-numbers.txt)
    \item \textbf{Single-core Energy Evaluation:} \X{} incurs average (maximum) energy overhead of 0.08\% (1.13\%) and 2.07\% (14.11\%) over the baseline with no RowHammer mitigation at \gls{nrh} = 1K, and at \gls{nrh} = 125, respectively. (plots/singlecore-energy-numbers.txt)
    \item \textbf{Single-core Performance Comparison:} \X{} outperforms Hydra, and PARA and performs similarly to Graphene, on average, at all \gls{nrh} values. \X{} incurs 0.08\% and 1.75\% performance overhead over Graphene at \gls{nrh}=1K and 125, respectively. \X{} outperforms Hydra by  0.67\% and 1.75\%, on average, at \gls{nrh}=1K and 125, respectively. \X{} outperforms Hydra by up to 39.19\% at \gls{nrh}=125. (plots/singlecore-comparison-numbers.txt)
    \item \textbf{Single-core DRAM Energy Comparison:}  \X{} exhibits lower energy consumption than Hydra, REGA, and PARA across all RowHammer thresholds. Second, CoMeT’s DRAM energy consumption is comparable to Graphene at all RowHammer thresholds. At \gls{nrh}=1K and 125, CoMeT incurs average energy overheads of 0.04\% and 0.96\%, compared to Graphene. Third, Hydra incurs more DRAM energy consumption over \X{} (by 0.39\% and 1.88\%. at \gls{nrh} = 1K and \gls{nrh} = 125, respectively). (plots/singlecore-energy-comparison-numbers.txt)
    \item \textbf{Sensitivity Analysis:} Please compare the figures.
    \item \textbf{Evaluating different reset period and \gls{npr} values:} Please compare the figures.
    \item \textbf{Motivational data:} Hydra has average (maximum) performance overheads of 0.85\% (8.18\%) at \gls{nrh} = 1K, and the average (maximum) performance overhead increases to 5.66\% (51.24\%) at a very low \gls{nrh} of 125 (plots/motiv-results.txt).
\end{itemize}

\subsection{Methodology}

Submission, reviewing and badging methodology:

\begin{itemize}
  \item \url{https://www.acm.org/publications/policies/artifact-review-and-badging-current}
  \item \url{http://cTuning.org/ae/submission-20201122.html}
  \item \url{http://cTuning.org/ae/reviewing-20201122.html}
\end{itemize}

\end{document}